\documentclass[10pt,conference]{IEEEtran}
\usepackage{caption}
\usepackage{subcaption}
\usepackage{braket}

\usepackage{todonotes} 

\usepackage{amsmath,amssymb,amsfonts}
\usepackage{algorithmic}
\usepackage{graphicx}
\usepackage{textcomp}
\usepackage{xcolor}
\usepackage{fancyhdr}

\usepackage[compact]{titlesec}

\titlespacing{\section}{0pt}{*0.5}{*0.5}
\titlespacing{\subsection}{0pt}{*0.05}{*0}
\titlespacing{\subsubsection}{0pt}{*0}{*0}

\usepackage[linewidth=1pt]{mdframed}
\usepackage{soul}
\usepackage{booktabs}

\usepackage{tikz}
\newcommand*\circled[1]{\tikz[baseline=(char.base)]{
            \node[shape=circle,draw,inner sep=0.5pt] (char) {#1};}}
\usepackage{mathtools}

\usepackage[font=bf,skip=0pt]{caption}
\usepackage{subcaption}
\usepackage{tabularx}

\ifCLASSINFOpdf
\else
\fi
\begin{document}

%
\title{Adaptive job and resource management for \\the growing quantum cloud}


\author{\IEEEauthorblockN{Gokul Subramanian Ravi}
\IEEEauthorblockA{University of Chicago\\
gravi@uchicago.edu}
\and
\IEEEauthorblockN{Kaitlin N. Smith}
\IEEEauthorblockA{University of Chicago\\
kns@uchicago.edu}
\and
\IEEEauthorblockN{Prakash Murali}
\IEEEauthorblockA{Princeton University\\
prakashmurali@gmail.com}
\and
\IEEEauthorblockN{Frederic T. Chong}
\IEEEauthorblockA{University of Chicago\\
chong@cs.uchicago.edu}

}


%


\maketitle
\thispagestyle{plain}
\pagestyle{plain}
\begin{abstract}
As the popularity of quantum computing continues to grow, efficient quantum machine access over the cloud is critical to both academic and industry researchers across the globe.
And as cloud quantum computing demands increase exponentially, the analysis of resource consumption and execution characteristics are key to efficient management of jobs and resources at both the vendor-end as well as the client-end.
While the analysis and optimization of job / resource consumption and management are popular in the classical HPC domain, it is severely lacking for more nascent technology like quantum computing.

This paper proposes optimized adaptive job scheduling to the quantum cloud taking note of primary characteristics such as queuing times and fidelity trends across machines, as well as other characteristics such as quality of service guarantees and machine calibration constraints. Key components of the proposal include a) a prediction model which predicts fidelity trends across machine based on compiled circuit features such as circuit depth and different forms of errors, as well as b) queuing time prediction for each machine based on execution time estimations.

Overall, this proposal is evaluated on simulated IBM machines across a diverse set of quantum applications and system loading scenarios, and is able to reduce wait times by over 3x and improve fidelity by over 40\% on specific usecases, when compared to traditional job schedulers.
\end{abstract}

\section{Introduction}
\label{Introduction}


Quantum computing is a revolutionary computational model that leverages quantum mechanical phenomena for solving intractable problems. 
Quantum computers (QCs) evaluate quantum circuits or programs in a manner similar to a classical computer, but quantum information's ability to leverage superposition, interference, and entanglement is projected to give QCs significant advantage in cryptography~\cite{Shor_1997}, chemistry~\cite{kandala2017hardware}, optimization~\cite{moll2018quantum}, and machine learning~\cite{biamonte2017quantum} applications.

In the current Noisy Intermediate-Scale Quantum (NISQ) era, we expect to operate with quantum machines comprising of hundreds or thousands of quantum bits (qubits), which are acted on by imperfect gates~\cite{preskill2018quantum}. 
Further, the connectivity in these machines will be sparse and qubits will have modest lifetimes. 
Given these limitations, NISQ era machines will be unable to execute large-scale quantum algorithms like Shor Factoring~\cite{Shor_1997} and Grover Search~\cite{Grover96afast}, which rely on error correction comprised of millions of qubits to create fault-tolerant quantum systems~\cite{O_Gorman_2017}.

With development of these NISQ devices, cloud-based quantum information processing (QIP) platforms with nearly 100 qubits are currently accessible to the public.
Further, recent quantum hardware roadmaps, such as IBM's~\cite{IBM-HW}, have announced that devices with as many as 1000 qubits will be available by 2023. 
It also has been recently demonstrated by the Quantum Supremacy experiment on the Sycamore quantum processor, a 54-qubit quantum computing device manufactured by Google, that quantum computers can outperform current classical supercomputers in certain computational tasks~\cite{arute2019quantum}. 
These developments suggest that the immediate future of quantum computing is promising.

\begin{figure}[t]
\includegraphics[width=\columnwidth,trim={0cm 0cm 0cm 0cm},clip]{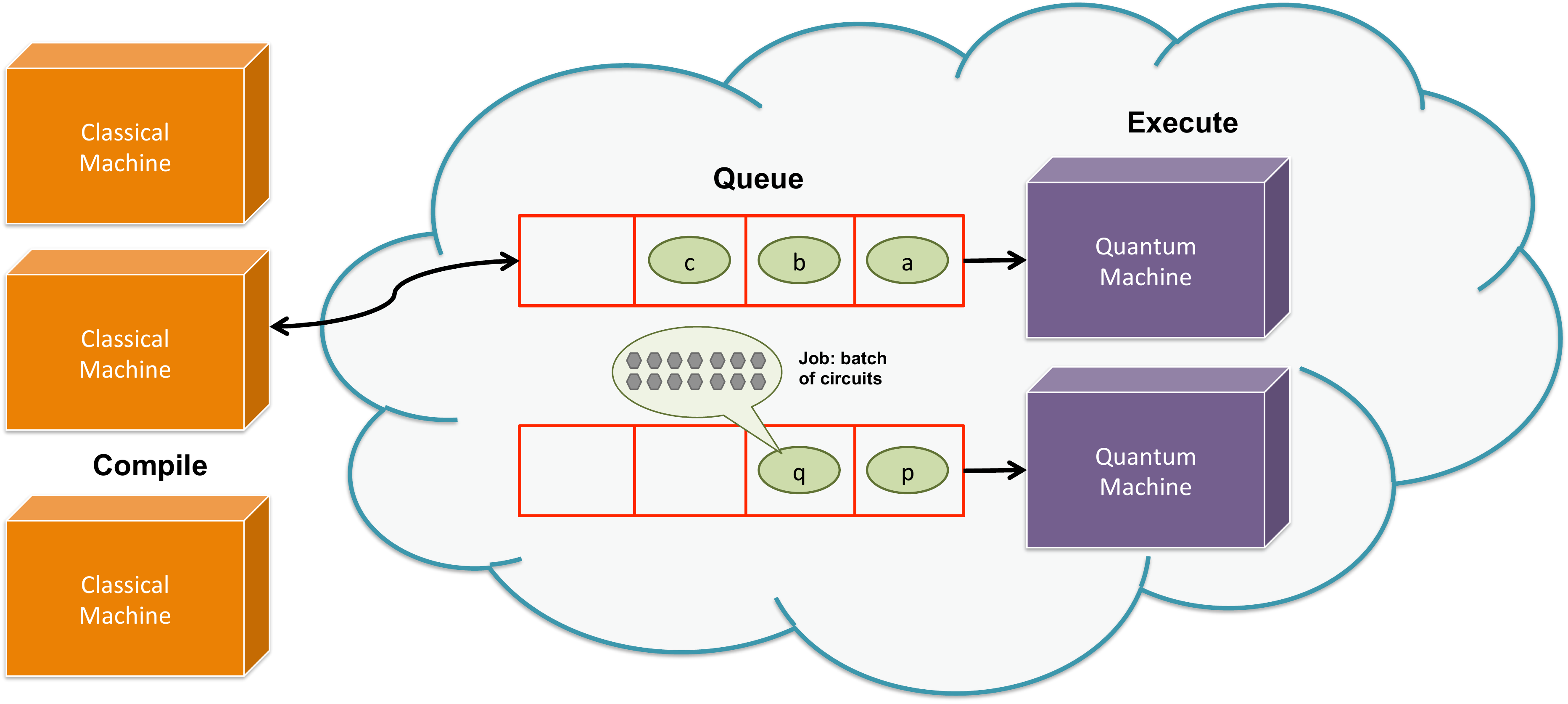}
\centering
\caption{Users launch quantum programs from their classical computers onto the vendor's quantum cloud wherein the jobs are queued until execution.} 
\label{Fig:SC21_Overview}
\end{figure}

Still at a nascent stage, QCs are an extremely scarce and expensive resource due to their difficulty to design, manufacture, and maintain.
Thus, quantum machines and corresponding software stacks are primarily accessed by researchers in academia and industry world wide via the cloud.
Current cloud vendors with their own quantum hardware include industry giants like IBM, Google, Microsoft and Honeywell, as well as startups such as Xanadu, Rigetti, IonQ and D-Wave.
Further, Amazon Braket (AWS) and Microsoft Azure Quantum provide quantum computing as a service via multiple other quantum hardware vendors.
It is expected that quantum computing as a cloud service will grow considerably over the next decade and will continue to be the main access to quantum machines for research across the globe.
Fig.\ref{Fig:SC21_Overview} provides an overview of how clients interact with cloud quantum machines - more details are discussed in Section \ref{background-motivation}.

Currently, the contention for access to quantum devices in the cloud is steadily growing. 
While quantum machines available in the cloud are very limited in number~\cite{IBMQE,AWS,Azure},  the number of users and the number of ``jobs” submitted to these machines are drastically growing every day~\cite{IBM-users} across multiple vendors. 
With the increasing popularity of quantum computing in both industry and academic research, it is expected that these contention trends will continue to worsen over the next decade or more - at the very least until the cost of building large and reliable quantum computers becomes more easily surmountable.
As an example, a first-order impact of quantum machine scarcity are the long queuing times~\cite{das:2019,kong2021origin} experienced while accessing cloud machines. 
As discussed in later sections, we observe that there can be 10s-1000s of quantum jobs queued up on quantum machines at any given time. 
This results in queuing times of many hours and sometimes even days.
Such accessibility constraints in using these machines can severely handicap several research endeavors in terms of: a) the scope of the quantum problems that can be effectively targeted on these machines, and b) timely access to the machines irrespective of the quantum problem.

%

Thus, as quantum demand continuous to grow, it is imperative to efficiently manage quantum resources.
Unfortunately, the current state of scheduling in the quantum cloud has not evolved significantly from the early days of quantum development where access to machines was more essential than optimum usage; when machines were very few in number, users were experts with advanced quantum knowledge and applications were very limited.
Today we have a vast diversity in user expertise, in target applications and in available machines meaning that optimal usage is essential from the perspectives of both the user as well as the vendor.
Thus, adaptive job and resource management suitable for the growing quantum cloud is in need.

Similar to classical HPC, vendors should attempt allocating machine resources as efficiently as possible so as to improve system throughput, while clients should try to make efficient use of job deployment strategies to maximize their allocated time and resources.
However there are some key differences between computing in the classical cloud and computing in the quantum cloud.
\circled{1}\ First, quantum machines are error prone. Thus maximizing execution fidelity is a first-order constraint unlike classical machines which are primarily focused on performance and energy efficiency.
\circled{2}\ Second, the execution of quantum applications are heavily dependent on, and sensitive to, the target quantum machine with its varying characteristics, meaning that any optimum scheduling in the cloud needs up to date machine information.
\circled{3}\ Third, in the near future quantum jobs / circuits are expected to be on the lower end of the complexity spectrum, meaning that their execution characteristics can be more easily predictable.

This paper proposes to automate, adapt and improve scheduling quantum jobs to the growing quantum cloud.
The improved scheduling targets a number of goals:
\circled{1}\ Maximizing execution fidelity at low system load, \circled{2}\ Minimizing wait times at high system load, \circled{3}\ A balanced approach otherwise, \circled{4}\ Accounting for user's QOS (Quality of Service) requirements, in terms of maximum acceptable wait times, \circled{5}\ Accounting for the effects of machine recalibration, and \circled{6}\ Optimizing calibration schedules for better overall fidelity / lower wait times.




\textbf{Overall, we make the following contributions:}
\begin{enumerate}
    \item This work shows, both qualitatively and quantitatively, that there is a need to improve the existing quantum job schedulers as the quantum cloud continues to grow.
    \item To the best of our knowledge, this is the first proposal to explore quantum job scheduling optimizations. 
    We build an automated adaptive job scheduler which can be integrated into the quantum cloud, to schedule quantum jobs onto machines, which optimizes for both fidelity and wait times, as well as accounts for the different objectives described earlier.
    \item We build a novel prediction model to predict correlation between compiled quantum circuit features and their machine execution fidelity, across a diverse set of quantum applications and quantum machines.
    \item We build a simple queuing time prediction model by estimating the execution times of jobs on quantum machines.
    \item We incorporate these prediction models into our proposed scheduler, and use them to balance different goals, meet QOS requirements and avoid stale compilation for machines.
    \item We further avoid stale machine compilation by exploring the relation between machine calibration cycles and job schedules and propose simple improvements to calibration schedules through the approach of ``staggering".
    \item We study the benefits of our proposal across a diverse set of quantum applications, a wide range of IBM quantum machines and different scenarios of system loading.
\end{enumerate}

\section{Background and Motivation}
\label{background-motivation}

\subsection{Traditional execution of quantum circuits}
We explain some key terminology in quantum circuit execution in the cloud below:

   \circled{1}\  \textbf{\emph{Circuit:}} A single quantum circuit with a list of instructions bound to some registers. It has a number of gates and is spread out over a number of qubits. 
    
    \circled{2}\  \textbf{\emph{Compilation:}} Involves a sequence of steps to enable the quantum circuit to be executed on a specified quantum machine in a valid and efficient manner.
    
    \circled{3}\  \textbf{\emph{Job:}} Encapsulates a single circuit or a batch of circuits that execute on an quantum machine. The circuits within a batched job are treated as a single task such that all quantum circuits are executed successively. Further, each circuit in the job will be rapidly re-executed for a specified number of shots. 
    
    \circled{4}\  \textbf{\emph{Queue:}} When a job is submitted to a quantum machine on the cloud, it enters a queue (for that particular machine) with jobs from other users before eventual execution. The order which these jobs are executed is, by default, determined by some fair sharing based queuing algorithm. 



\subsection{Long queuing times in today's quantum systems}

\begin{figure}[t]
\includegraphics[width=\columnwidth,trim={0cm 0cm 0cm 0cm},clip]{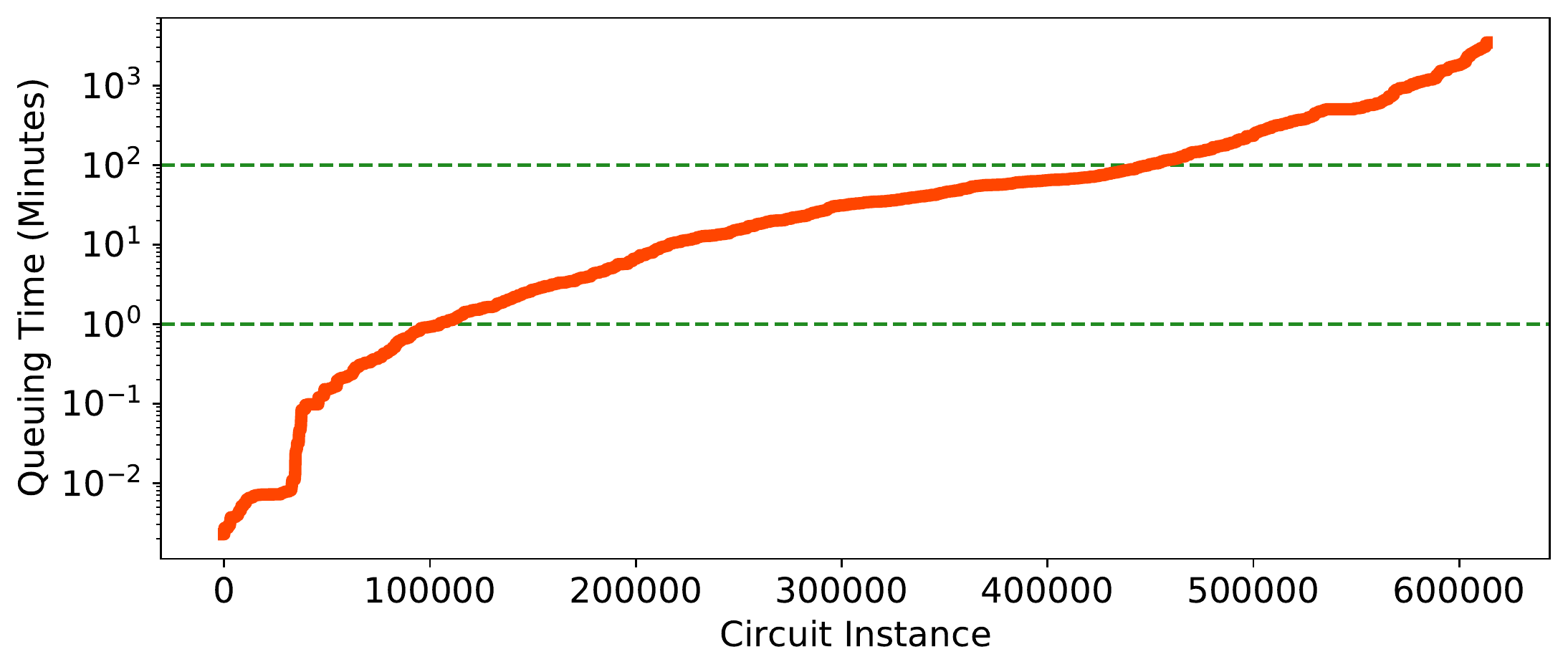}
\centering
\caption{Queuing time experienced by circuits run on the IBM Quantum machines (sorted) over two years. Green lines correspond to times of 1 minute and 2 hours.} 
\label{Fig:SC21_QT}
\end{figure}

Long queuing times experienced while accessing cloud machines come as a direct result of quantum machine scarcity. 

Fig.~\ref{Fig:SC21_QT} plots the cloud queuing time experienced by the executed circuits in an in-house study we conducted over a two year period, in ascending order.
Note that these executed circuits are through a mix of public and privileged (i.e. paid) access to these quantum machines.
Only around 20\% of the total circuits experience ideal queuing times of, say, less than a minute.
The median queuing time is around 60 minutes which is not insignificant.
Further, more than 30\% of the jobs experienced queuing times of greater than 2 hours, and around 10\% of the jobs were queued up for as long as a day or even longer!
The classical HPC systems analyzed in \cite{Patel:2020} estimated that the average queuing times on their supercomputers increase from 0.1 hours to 1.2 hours over a decade.
The current queuing times for quantum clouds, even at this stage of relative infancy, are already comparable to the higher side of the classical queuing times.
A similar 10x increase in quantum waiting times over the next decade would be detrimental to quantum research and development.
The higher queuing times are especially concerning, considering that the actual quantum execution runtime on the quantum machines is only in the order of seconds or minutes.

\emph{\textbf{Takeaway:} Queuing times are considerably long and are expected to grow as demand increases. It is important that jobs are scheduled in a more queuing time aware manner.}

\subsection{Queue time variability across machines}

\begin{figure}[t]
\includegraphics[width=\columnwidth,trim={0cm 0cm 0cm 0cm},clip]{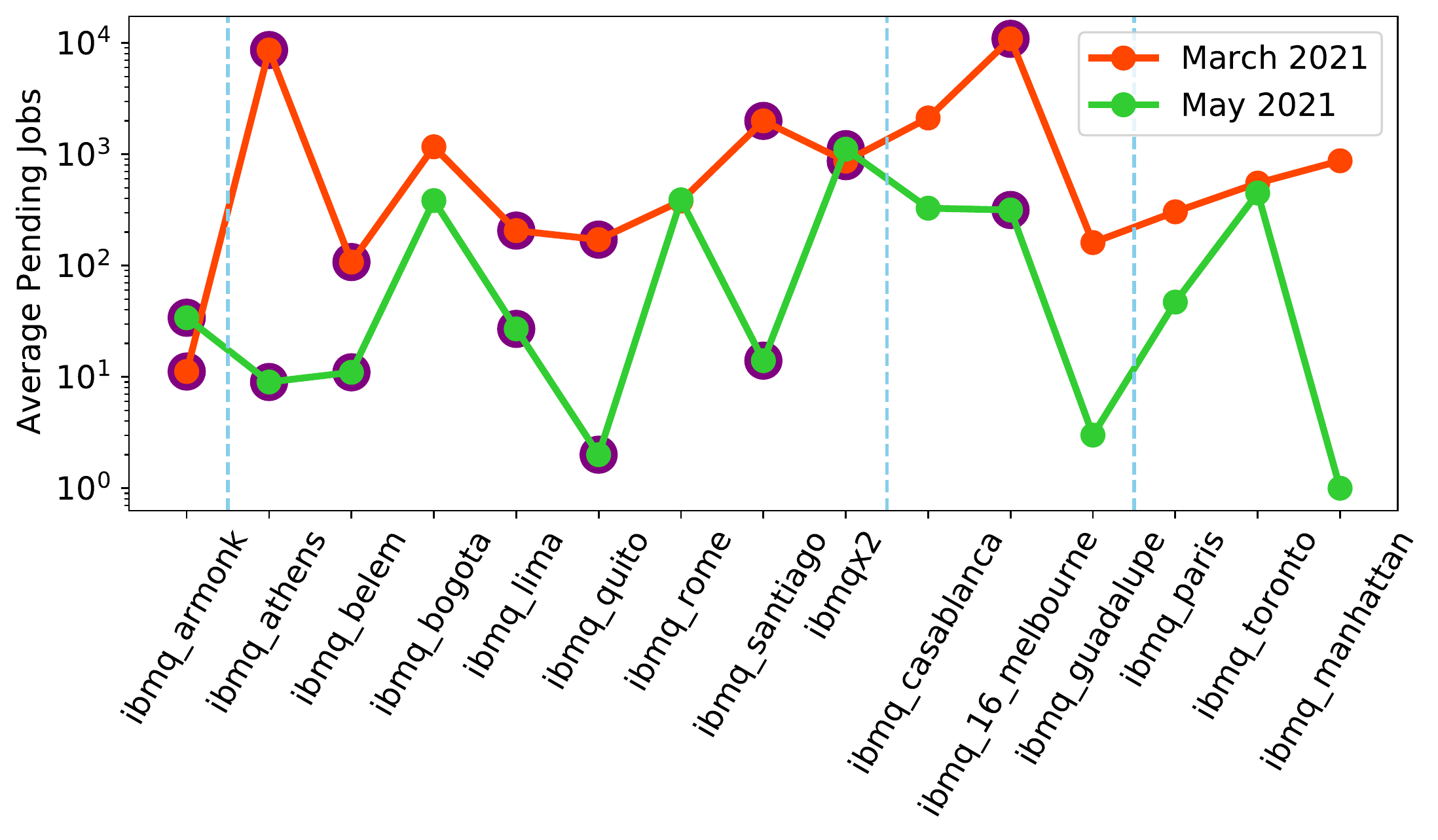}
\centering
\caption{Average pending jobs across different quantum machines, averaged over a week in March and May 2021} 
\label{Fig:SC21_MachineQJ}
\end{figure}

Fig.~\ref{Fig:SC21_MachineQJ} shows the number of pending jobs across different quantum machines, averaged over a week's period in March and May 2021.
The machines are broken down into blocks (blue dashed lines) based on the number of qubits in the machine.
The first block is a 1-qubit machine, the next block is 5-qubit machines, the next is 7-16 qubits and the final is 27-65 qubits.
Further, publicly accessible machines are highlighted in purple.
In each block (and across both periods), it is observed that the average pending jobs are highest on a public machine - this is expected since public machines have considerably more demand.
For instance, in the March week, IBMQ Athens is 10-100x more in demand than other 5-qubit machines.
More importantly, it is observable that jobs are not distributed equally across machines (public or otherwise).
And further, trends in job distribution are not stable over time - distributions vary widely between the two weeks shown.
It is intuitive that while specifics of machine usage and machine popularity might change over time, the trends that jobs are unequally distributed across machines and that public machines are considerably in higher demand are expected to be consistent. 

\emph{\textbf{Takeaway:} Jobs are unequally distributed across quantum machines. It is important to ensure that load is well balanced across the system in order to reduce unreasonable queuing times on specific machines - especially in the public cloud where demand is considerably higher.}

\begin{figure}[t]
\includegraphics[width=\columnwidth,trim={0cm 0cm 0cm 0cm},clip]{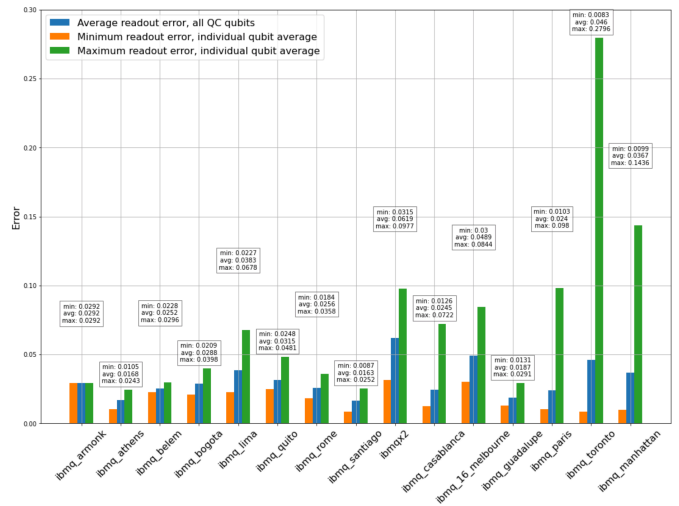}
\centering
\caption{Variation in readout error over 74 calibration cycles across IBM machines.} 
\label{Fig:QCE21_ro}
\end{figure}

\begin{figure}[t]
\includegraphics[width=\columnwidth,trim={0cm 0cm 0cm 0cm},clip]{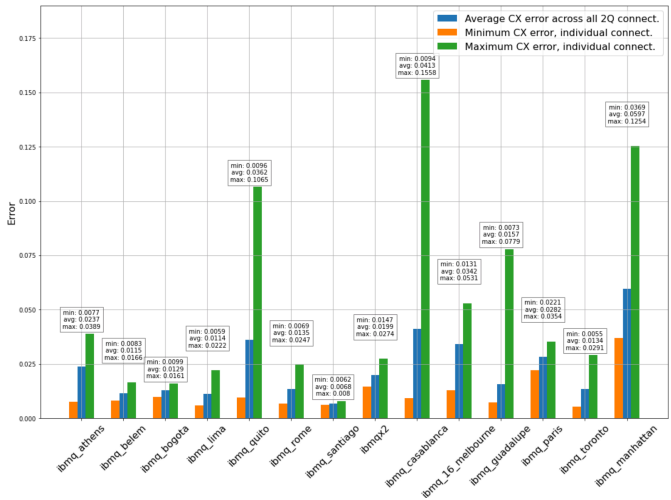}
\centering
\caption{Variation in CX error over 74 calibration cycles across IBM machines.} 
\label{Fig:QCE21_cx}
\end{figure}

\subsection{Spatial variability across machines}

Even if QCs are manufactured in a highly controlled setting, unavoidable variation results in intrinsic properties that impact performance. This variation between devices becomes especially apparent when examining error rates. In Fig.~\ref{Fig:QCE21_ro}, details of average qubit readout error for 15 IBM QCs over 74 calibration cycles from April 2021 to June 2021 are plotted. 
Here, minimum and maximum per-qubit averages over the 74 cycles are included along with the overall average readout error per machine. The QCs are in ascending order from left to right in with respect to of number of qubits.
Even if machines have the same number of qubits, such as the five qubit devices ranging from ibmq{\_}athens to ibmq{\_}santiago on the x-axis, their readout error values differ. 
Understanding a QC's readout error is important as qubit measurement is a required step in QIP. 

Fig.~\ref{Fig:QCE21_cx} plots average CNOT (CX) error for 15 IBM QCs in asending order in terms of number of machine qubits over 74 calibration cycles from April 2021 to June 2021. Here, the minimum and maximum average individual CX operation over 74 cycles is included along with total average CX error per machine. Once again, the data show that each machine differs in two-qubit gate perfomance. Additionally, the spread between intra-machine minimum and maximum CX error is not insignificant.

\emph{\textbf{Takeaway:} Machine characteristics can vary widely across machines. Moreover, many characteristics and their impact on applications are likely not be understood well by users. Thus analyzing how different machine characteristics affect application fidelity and scheduling jobs to machines accordingly are an important step in effective quantum job scheduling.}


\begin{figure}[t]
\includegraphics[width=\columnwidth,trim={0cm 0cm 0cm 0cm},clip]{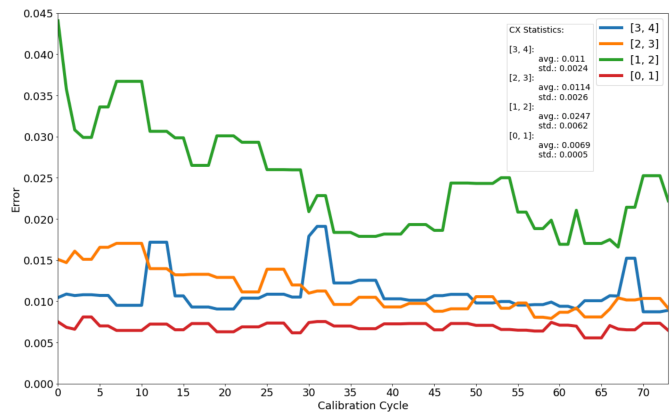}
\centering
\caption{Variation in CX errors over time (74 calibration cycles) on IBM Rome. Each line represents a two-qubit connection on the QC that executes a CX gate.} 
\label{Fig:QCE21_rome_cx}
\end{figure}

\subsection{Temporal variability across calibrations}

When quantum circuits are compiled, they are done so in a device aware manner. 
While this involves static characteristics such as device topology and device basis gates, it also involves incorporating dynamic characteristics such as gate / qubit fidelity. 
As discussed earlier, the latter are dynamic because they evolve over time - these characteristics of qubits and gates are re-calibrated at some coarser granularity (say once a day) and these calibrations are non-uniform i.e. one day's qubit fidelity can be very different from the next day's qubit fidelity. 
Further, these characteristics also drift over time - meaning that they can differ even within a single calibrated epoch. 

To showcase variation among calibration cycles, Fig.~\ref{Fig:QCE21_rome_cx} illustrates how drastically CX error changes over time on a QC. Here, the CX error rates for ibmq{\_}rome are plotted for 74 calibration cycles ranging from April 2021 to June 2021. Data describing the average and standard deviation for each QC CX gate over the sample period is also included. As seen in Fig.~\ref{Fig:QCE21_rome_cx}, CX operations have unique error characteristics, some of which, like CX [1, 2], that evolves significantly over time.

\begin{figure}[t]
     \centering
     \begin{subfigure}[b]{0.48\columnwidth}
         \centering
         \includegraphics[width=\textwidth]{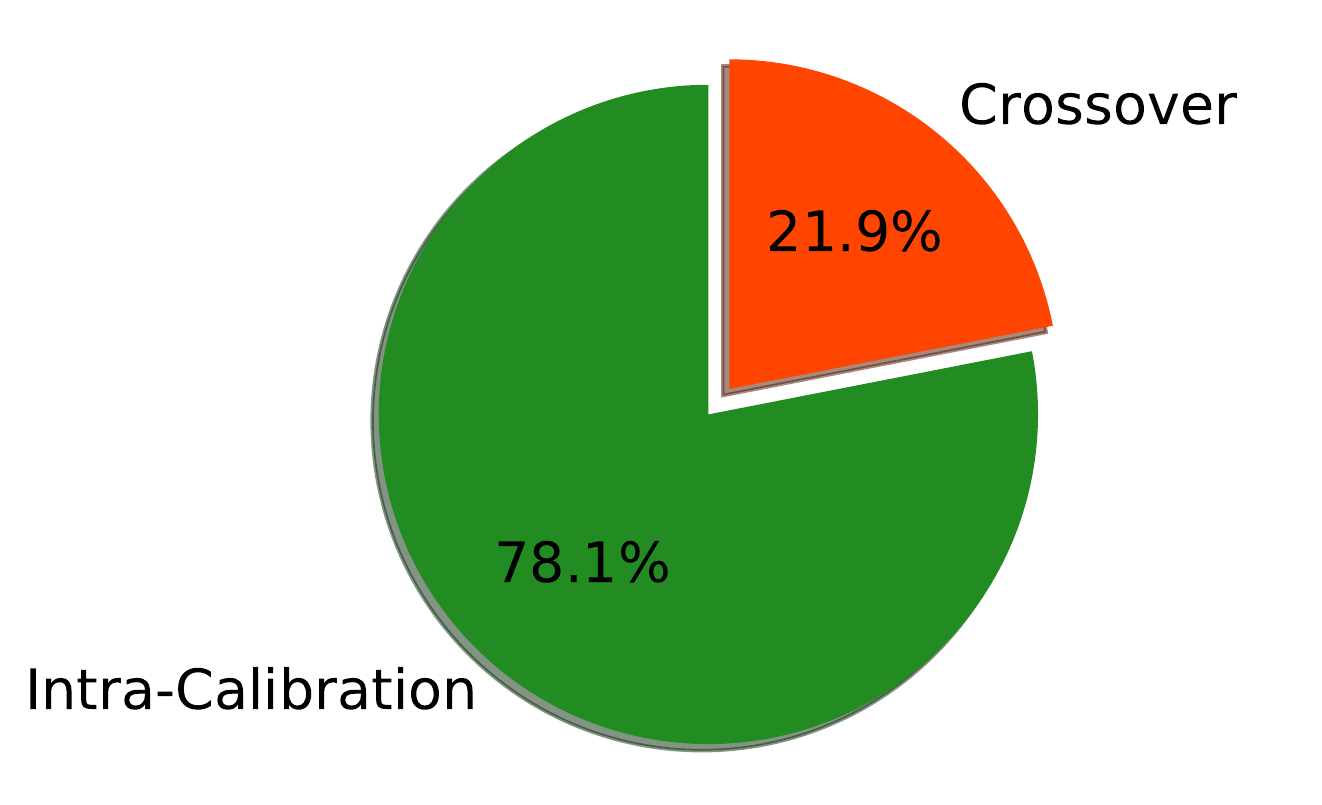}
         \caption{Jobs crossing calibrations}
         \label{Fig:SC21_Cross1}
     \end{subfigure}
     \begin{subfigure}[b]{0.5\columnwidth}
         \centering
         \includegraphics[width=\textwidth]{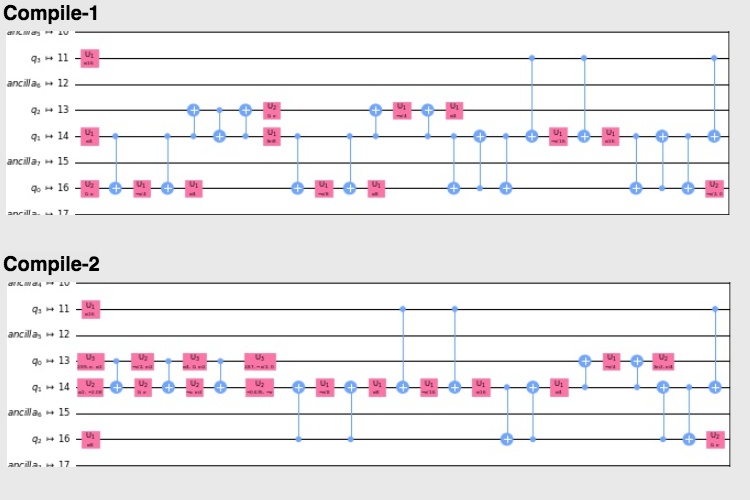}
         \caption{Varying Compiled Circuits}
         \label{Fig:SC21_Cross2}
     \end{subfigure}
        \caption{Effects of calibration crossovers. QCs can become sub-optimal over time.}
\end{figure}

It is also often the case that in scenarios of long queuing times, the dynamic characteristics which are accounted for at the  time of compilation are very different from the dynamic characteristics of the quantum machines at the time when the quantum circuit is actually executed on the machine. 
This results in the quantum circuit being sub-optimal to the quantum machine at the time of actual execution.

IBM Quantum machines are usually calibrated once a day, likely around 12:00am - 2:00am (North America).
Fig.\ref{Fig:SC21_Cross1} shows that we estimate that over 20\% of our studied quantum jobs were compiled with device information from an older calibration cycle but were executed on the machine after a new calibration.
This results in the compilation being potentially sub-optimal.
Note that these are only coarse estimates based on queuing and execution time stamps.

Fig.\ref{Fig:SC21_Cross2} shows a snippet of a circuit compiled with noise-aware mapping, wherein the noise information of physical qubits is incorporated into the optimal mapping from the circuit's logical qubits to the machine's physical qubits.
Two compilations of the same circuit snippet are shown, from two consecutive calibration cycles.
It is evident that the optimal mapping and circuit structure are different.
Thus, using an older mapping can be detrimental to the fidelity of executed applications.

Note that in the above gate-based compilation approach, the quantum gates are converted to pulses at the time of execution. 
Thus the system will presumably use the most-recently-calibrated pulses to execute the gates on the quantum machine i.e. after the job reaches the head of the queue and is ready for actual quantum execution. 
On the other hand, in the pulse based approach (eg. OpenPulse~\cite{pulse-alexander2020qiskit,pulse-mckay2018qiskit,pulse-gokhale2020optimized}), pulses are generated at the time of compilation.
Thus, these pulses are generated based on machine characteristics at the time of compilation.
A calibration cross-over would mean that even the pulses are sub-optimal at the time of quantum execution.

\emph{\textbf{Takeaway:} Machine characteristics vary constantly, especially after day-to-day calibration. Thus time-sensitive machine characteristics have to be accounted for during scheduling. Moreover, scheduling strategies should be designed to avoid the machine execution of circuits compiled with stale machine data. }

\subsection{Classical scheduling in the cloud}
An increasing amount of computing is hosted in public clouds, such as those from Amazon, Microsoft, Google etc. 
Cloud platforms provide two major advantages for end-users and cloud operators: flexibility and cost efficiency. 

Optimized job scheduling and resource management in the cloud has been an active area of research and development over the past two decades including optimizations targeting resource reservation vs sharing~\cite{Cloud1}, QOS-aware scheduling~\cite{Cloud2} and maximizing system utilization~\cite{Cloud3}, to name a few.
Incorporating fairness and priorities in scheduling is not new to the quantum cloud.
For example, IBM Quantum applies a "fair share" approach to job scheduling~\cite{IBM-Fair}.
In this approach, jobs on a quantum system are managed dynamically  so that no user / group can monopolize the system. 
The shares provided to each user / group represent the fraction of system time that is allocated to them. 
Those with the most device time have the highest priority in the fair-share algorithm. 

Optimized job scheduling to the quantum cloud can differ from its classical counterpart in at least two distinct ways.
One, the execution of quantum applications on the quantum machine are significantly impacted by machine fidelity characteristics (such as static qubit connectivity within the machines and dynamic qubit error rates) as discussed prior.
This adds an addition layer of "heterogeneity" to the optimum job schedule.
Second, in the near future, quantum jobs / circuits are expected to be at the lower end of the complexity spectrum, meaning that both the execution time (which can then be extrapolated to queuing time) as well as the machine-application fidelity correlation can be reasonably predicted.
This implies that opportunities for job scheduling that optimize for both fidelity and wait times, as well as a multitude of other user/system requirements, are worthy of exploration.

\emph{\textbf{Takeaway:} As the quantum cloud matures, it is imperative to employ job scheduling strategies inspired by those pursued in the classical computing domain. However, key differences form the classical domain should be accounted for, especially those related to fidelity and execution characteristics.}
\begin{figure*}[t]
\includegraphics[width=0.85\textwidth,trim={0cm 0cm 0cm 0cm},clip]{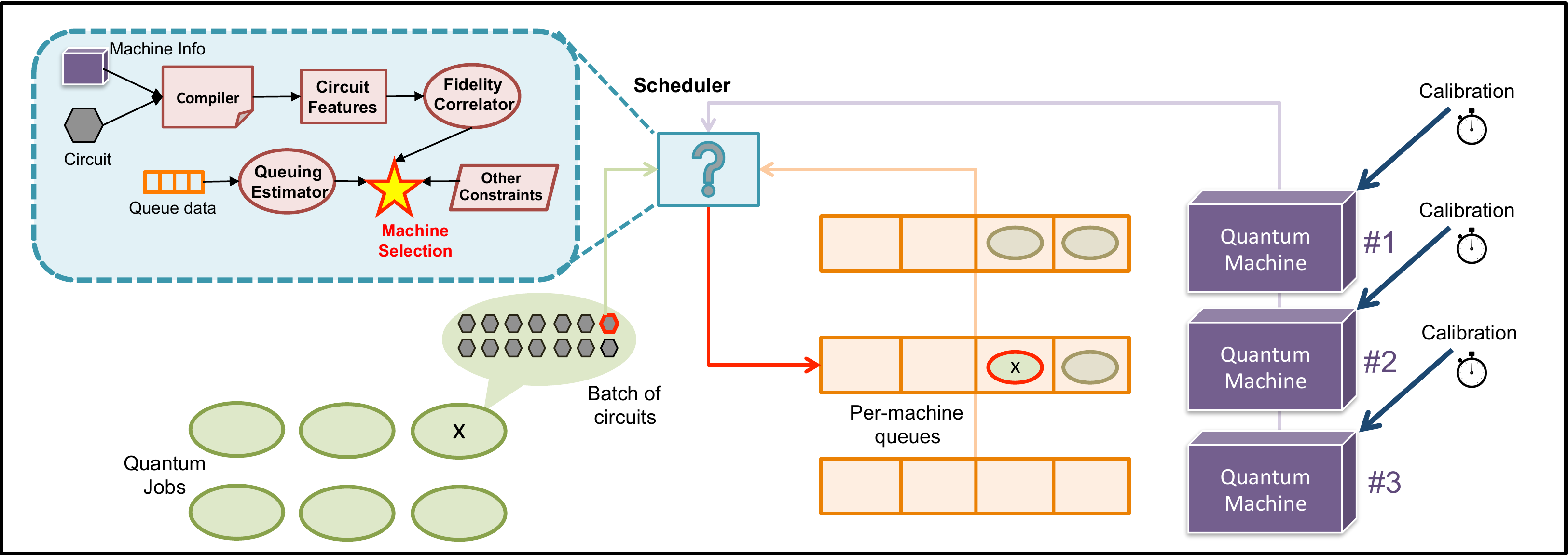}
\centering
\caption{Jobs are scheduled onto an appropriate quantum machine by the job scheduler. The scheduler compiles the quantum circuit i for different machines and extracts some key features. The features are then correlated with the fidelity of the machine using a novel predictor. In parallel, queuing and job data are used to estimate waiting time on each machine. Based on this information and other constraints, an appropriate machine is selected for the job.} 
\label{Fig:QCE_Prop}
\end{figure*}

\section{Experimental Setup}

\subsection{Compilation}
Applications are compiled with highest optimization offered by Qiskit Terra to map and optimize for the IBM machines~\cite{Qiskit}. 
We use Qiskit's noise aware compilation strategies~\cite{murali2019noise} to use less noisy qubits most efficiently and thus maximizing the likelihood of successful runs.
The above maps a circuit to a particular machine using the daily calibration data provided by the  vendor in order to avoid using unreliable qubits and to prioritize qubit positioning which reduces the likelihood of communication (SWAP) errors. 

\subsection{Quantum cloud simulation infrastructure}
For our experimental evaluation we build a model cloud setup using IBM Qiskit~\cite{Qiskit} simulator with device models of IBM quantum machines. 
We utilize device models of as many as 26 IBM quantum machines - these models mimic the characteristics of the actual device in terms of topology, error rates for gates and readout, T1/T2 times etc.
Machine details can be found on the IBM Quantum Systems page~\cite{IBMQS}.
Thus fidelity estimations are obtained by running the quantum circuits on these simulator models.

To model the system load we build different load distributions of low, high and random queuing jobs / times across these machines. 
Loads are defined with respect to a maximum queuing time which cannot be overshot. 
For example, given that machines are traditionally recalibrated every 24 hours, the max queuing time could be set to 24 hours.
\emph{Low load:} Less than 10\% of maximum queuing on each machine, \emph{High load:} 50-100\% of the maximum queuing on each machine, \emph{Random Load:} anywhere from 1-100\% of the maximum queuing on each machine.


\subsection{Benchmarks}
The framework is evaluated on benchmarks representative of real-world use cases, which are described below. 

\emph{Toffoli:} 
A 3-input gate which performs logical AND between two controls bits and writes onto the target bit. 

\emph{Hidden Subgroup Problem:} 
Captures problems like factoring, discrete logarithm, graph isomorphism, and the shortest vector problem. It is implemented for 4 qubits.

\emph{Bernstein-Vazirani:}
BV guarantees the return of the bitwise product of some input with a hidden string~\cite{bernstein1997quantum}. 
BV is implemented using 5 qubits.     

\emph{Linear Solver:} Solver for a linear equation utilizing 3 qubits.

\emph{Quantum Approximate Optimization Algorithm:} 
QAOA~\cite{farhi2014quantum} is implemented atop a parameterized circuit called an ansatz and we use one instance of a hardware efficient QAOA ansatz as the benchmark.
We use QAOA ansatz for 4 qubits.

\emph{Variational Quantum Eigensolver:}
The goal of this algorithm~\cite{peruzzo2014variational} is to variationally find the lowest eigenvalue of a given problem matrix.
We implement VQE on a hardware-efficient SU2 ansatz~\cite{IBM-SU2} and use one instance as the benchmark.
We construct the ansatz for 4 qubits (4 reps / full entanglement) and 6 qubits (3 / SCA).

\emph{Quantum Repetition Code Encoder:}
A repetition code encoder which introduces redundancy to the encoding that can be exploited for error detection~\cite{Roffe_2019} (5 qubits).

\emph{Ripple Carry Adder:}
We implemented a linear-depth, 2 bit ripple-carry adder quantum circuit that uses 6 qubits based on the structure described in~\cite{cuccaro2004new}.

\subsection{Metrics}
The benefits of our scheduler are primarily evaluated for a) fidelity, based on the \emph{Probability of Success (POS)} metric which is the ratio of a number of error-free trials to the total number of trials, and b) queuing time, based on a simulated load distributions  across our model cloud system made up of ``fake" machines (i.e. simulation).

\subsection{Evaluation comparisons}
We compare our proposed scheduler against two baselines - \emph{Only-WT:} which only aims to minimize wait times (i.e. queuing times) agnostic to application fidelity or QOS requirements etc, and \emph{Only-Fid:} which only aims to maximize application fidelity (based on predictions from the fidelity correlator) but is agnostic to system load, wait times etc.

\section{Proposal}
This paper proposes to automate and improve the scheduling of quantum jobs into the quantum cloud.
The improved scheduling targets a number of goals:
\circled{1}\ Maximizing execution fidelity at low system load, \circled{2}\ Minimizing wait times at high system load, \circled{3}\ A balanced approach otherwise, \circled{4}\ Accounting for user's QOS (Quality of Service) requirements, in terms of maximum wait times, \circled{5}\ Accounting for the effects of machine recalibration, and \circled{6}\ Optimizing calibration schedules for better overall fidelity / throughput.

An overview of the proposal is shown in Figure \ref{Fig:QCE_Prop} and summarized below. Detailed design evaluation in Section \ref{sec_DE}:

\circled{1}\ A user uses their classical computing device to launch a quantum job to be executed in the cloud. Unlike the traditional setting, the user does not specify a target quantum machine in the vendor cloud. The machine selection will be managed by the scheduler.

\circled{2}\ A job's QC is compiled for all suitable machines. Unsuitable machines can include those that have lower number of qubits than the circuit requires, queuing times beyond the QOS specification of the application etc.

\circled{3}\ Note that at the above step, in case a batch of circuits are present in the job, the user is allowed to specify a specific circuit to be representative of the batch - to ease compilation overheads across a variety of machines.

\circled{4}\ Once the circuit is compiled for the suitable machines, post-compilation features of the circuit for each machine are extracted and passed to the fidelity correlator.

\circled{5}\  The fidelity correlator provides a correlation between the circuit features and the expected fidelity of the execution of the quantum circuit on each machine.

\circled{6}\  In parallel, the job queuing information on each machine, along with the sizes of the jobs and the number of shots of execution, are used to predicting the wait times on each machine.

\circled{7}\  Other constraints such as QOS requirements, machine calibration information are taken into account.

\circled{8}\  A machine selection is made based on all of the above information using a utility function. The utility function is built to optimize for fidelity and wait times, as well as to respect other constraints such as QOS and calibration.

\circled{9}\ Once the machine is selected, any uncompiled circuits in the job (which were not used for machine selection) are compiled for the target machine.

\circled{10}\ Finally, the job joins the machines queue and waits for execution (note that in a more optimized design, it could be possible to overlap the last two steps).

\circled{11}\ Further, the scheduler can provide inputs to space out the recalibrations of machines so as to better maximize the system efficiency.

\section{Design and Evaluation}
\label{sec_DE}

In this section we evaluate job scheduling policies based on fidelity, exeuction times and system load.
Fidelity is evaluated via simulated IBM quantum machines which are a snapshot representation of the actual machine.
Execution times are evaluated from data collected over millions of circuits run on the machines themselves over a two year period.
Machine load is simulated via an in-house queuing model model which interacts with the above. 

\begin{figure}[t]
\includegraphics[width=\columnwidth,trim={0cm 0cm 0cm 0cm},clip]{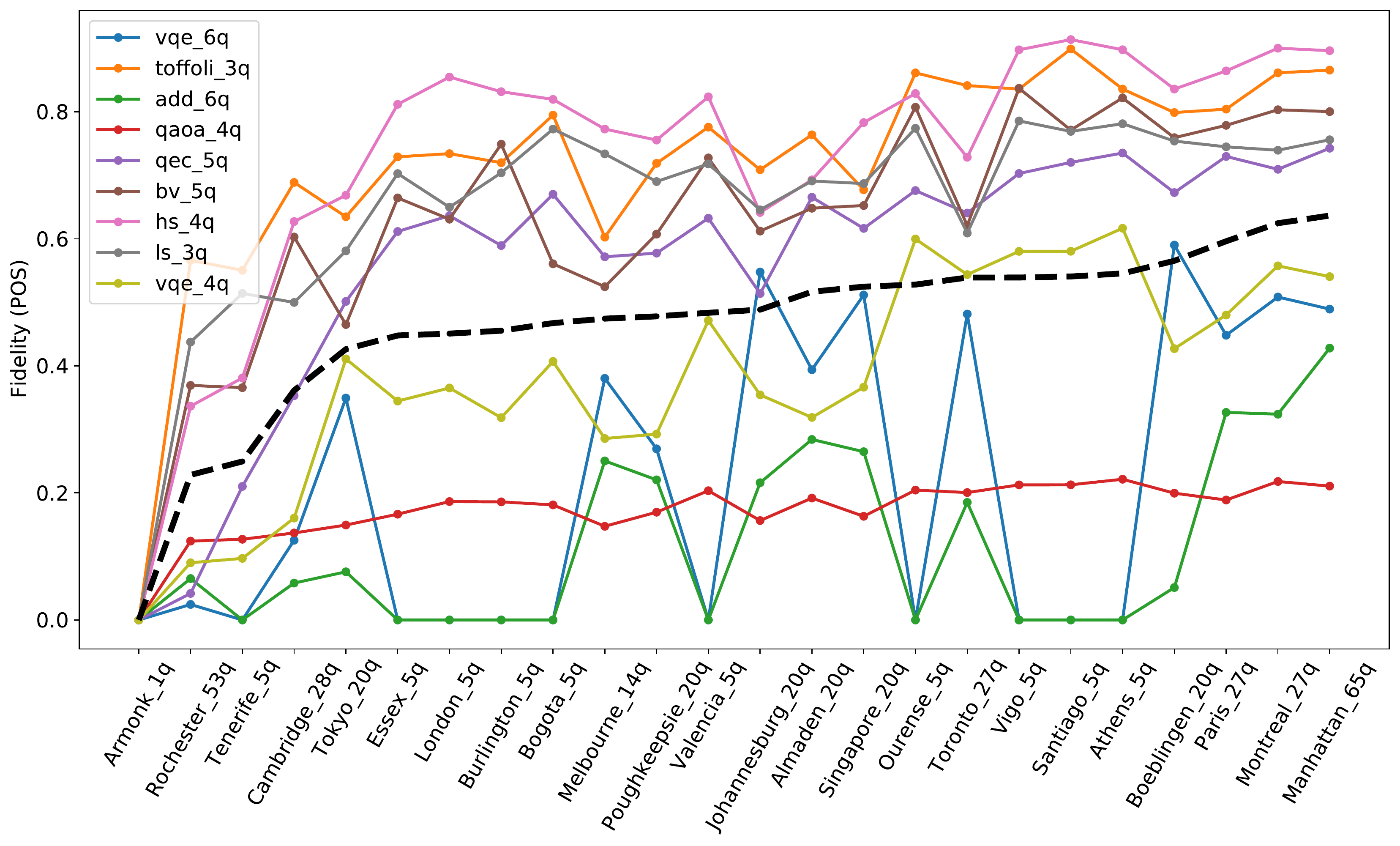}
\centering
\caption{Fidelity across (simulated) quantum machines.} 
\label{Fig:QCE_Machine_Fid}
\end{figure}

\subsection{Predicting the best machine (fidelity) for the job}
Fig.\ref{Fig:QCE_Machine_Fid} shows the fidelity of 9 benchmarks on the 26 simulated quantum machines. 
The dashed line shows the average fidelity on these machines and machines are sorted by this average.
It is evident from the figure that fidelity trends exist - some machines such as Athens - Manhattan consistently perform better than other machines.
Note that the correlation isn't purely related to the size of the machines.
While the largest machine (Manhattan - 65q), and larger machines (Paris - 27q etc) are on the right end of the graph, so are machines like Athens and Santiago which are 5q machines.
Further, even among these machines trends are not always fixed.
For example, the 5q machines sometimes outperform machines such as Manhattan, Paris etc (i.e. all the lines are not monotonically increasing).
Thus it is clear that there are potential macroscopic trends within machine behavior but they are not simple enough to be naively captured.

\begin{figure}[t]
\centering
\includegraphics[width=0.85\columnwidth,trim={0cm 0cm 0cm 0cm}]{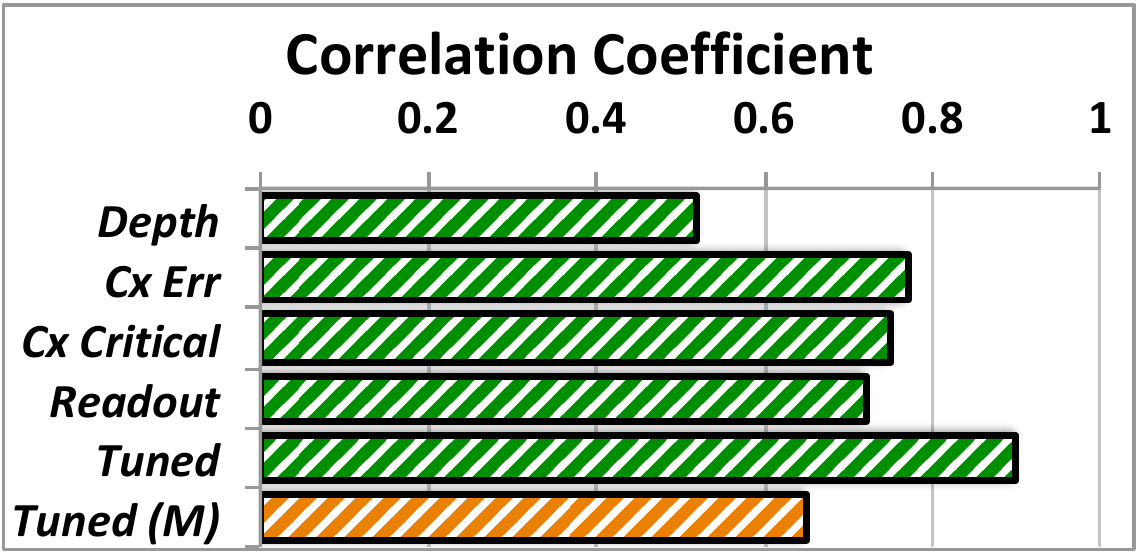}
\caption{Correlation between predictions and actual fidelity results. 0.5-0.7 is moderate, 0.7-0.9 is highly correlated.}
\label{fig:machine_corr}
\end{figure}

To build a fidelity correlator (as shown in Fig.\ref{Fig:QCE_Prop}), we make use of four features that are characteristics of a circuit compiled to a particular quantum machine and which intuitively affect the fidelity of the circuit when run on the machine.
These features are: a) Circuit Depth, b) Avg. CX error over the circuit, c) Avg CX in the circuit critical path, and d) readout errors on the measured qubits.
Note that there are other features (eg. 1q gate errors) which can also be considered to improve the quality of this metric.

The model is built as a product of linear terms: $F_n = \Pi(a_i + b_i*x_i)$, where $F_n$ is the fidelity of job $n$, $x_i$ is the feature and $a_i$ and $b_i$ are the tuned coefficients.
The model is developed with \emph{scipy.optimize curve\_fit}.
Collected data is split into training and test sets (70 / 30\%) to build the model (we also validated the results on a 33\% training set).

In Fig.\ref{fig:machine_corr} the Pearson correlation between actual application fidelity and the tuned model ("Tuned"), as well as with each feature is shown.
Bars in green show results averaged over the 26 simulated machines.
The orange bar shows results averaged from 15 real quantum machines run on the cloud.
Correlations in the range of 0.5-0.7 are considered moderately correlated while correlation greater than 0.7 is considered highly correlated.
First, note that in simulation all the features show moderate correlation against the application fidelity.
The tuned model shows very high correlation, achieving a coefficient of nearly 0.9.
On the real machines, the tuned model "Tuned (M)" achieves a correlation of near 0.7 which is at the borderline of moderate and high correlation.
Thus, it is clear that even a simple model with a few features is able to capture fidelity correlation with moderate to high accuracy.
Higher accuracy can potentially be achieved by adding more features as well as improving the model itself.

\begin{figure*}[t]
     \centering
     \begin{subfigure}[b]{0.73\textwidth}
         \centering
         \includegraphics[width=\textwidth,trim={0cm 0.5cm 0cm 0cm},clip]{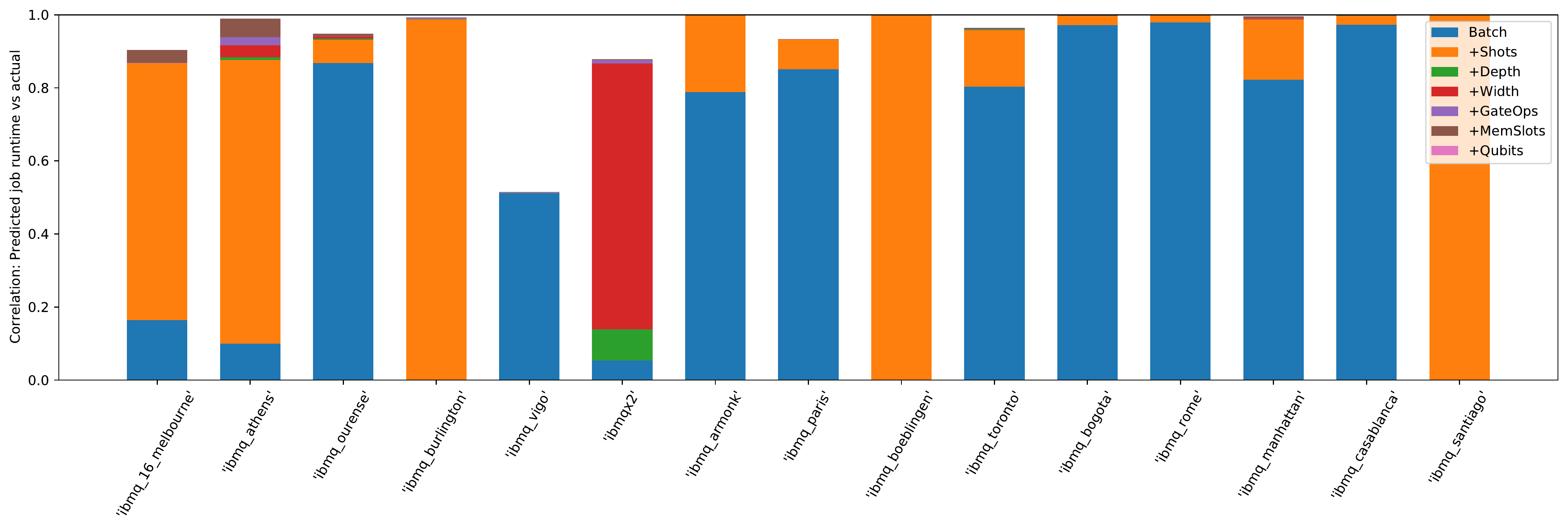}
         \caption{Job characteristics vs runtimes}
     \end{subfigure}
     \begin{subfigure}[b]{0.24\textwidth}
         \centering
         \includegraphics[width=\textwidth]{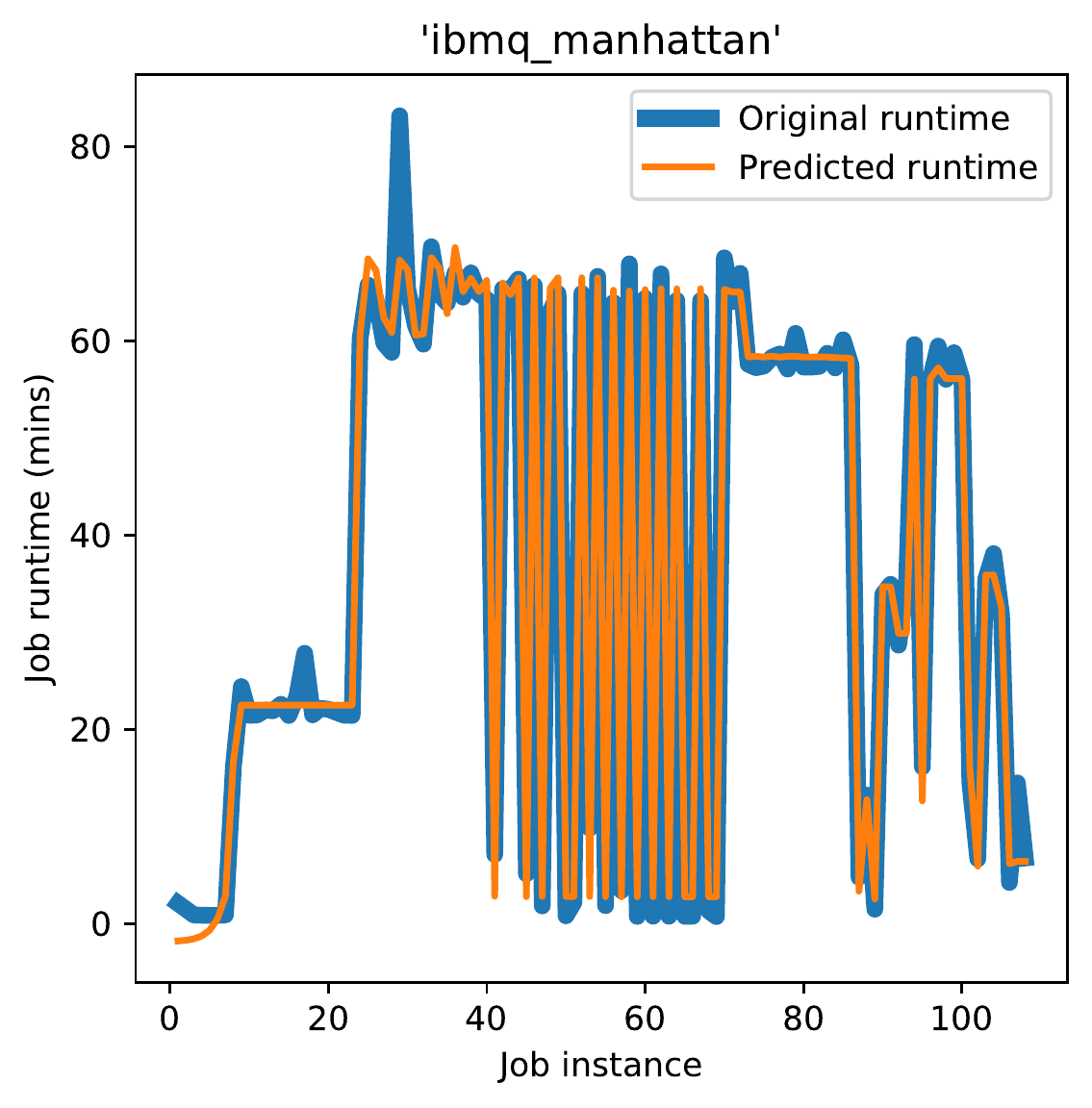}
         \caption{Predicted vs. Actual }
     \end{subfigure}
        \caption{Correlating the predicting runtimes (based on job characteristics) with actual observed runtimes. The major contributor to the correlation is the batch size.
A second contributor is the number of shots.}
        \label{Fig:QCE21_ExTime}
\end{figure*}

\subsection{Predicting execution times and, thus, queuing times}
\label{ExePred}
To understand the dependencies of execution time on job characteristics,  we build another simple prediction model.
The model is built as a product of linear terms: $E_n = \Pi(a_i + b_i*x_i)$, where $E_n$ is the execution time of job $n$, $x_i$ is the feature and $a_i$ and $b_i$ are the tuned coefficients.
The studied features are: batch size, number of shots; circuit: depth, width  and total quantum gates; and machine overheads: size (proportional to qubits) and memory slots required.
The model is developed with \emph{scipy.optimize curve\_fit}.
Collected data is split into training / test sets (70 / 30\%) to build the model.

Fig.\ref{Fig:QCE21_ExTime}.a plots the correlation of predicted runtimes vs actual runtimes, averaged across all jobs that ran on each quantum machine.
Correlation is calculated with the Pearson Coefficient.
First, note that the correlation is 0.95 or above on all but two machines.
The major contributor to the correlation is the batch size, i.e. the number of circuits in the job.
A second contributor is the number of shots which is usually influential when the batch size of the job is low.
Other factors like depth, width and memory slots have limited influence -  suggesting that batching and shots are the main contributors.

In Fig.\ref{Fig:QCE21_ExTime}.b we plot the actual runtimes for different jobs on a particular machine, IBMQ Manhattan in comparison to the predicted runtimes.
The high accuracy in prediction is evident.

Thus, while machine and job characteristics can vary widely, application's runtimes remain fairly predictable.
This is primarily because we are in the early stage of quantum computing exploration in which the number of qubits are low and the algorithmic depth and complexity of the circuits are limited.
Therefore the overheads associated with execution of a circuit is more influential than the characteristics of the circuit itself - this trend is expected to persist in the near term.

Finally, summing up the predictions of execution times across a machine's queue, provides an estimate of its queuing time as follows: $Q_M = \Sigma^{m} E_i$, where $Q_M$ is the queuing time on machine $M$, $E_i$ is the execution time of the ith job in machine $M$'s queue which has a total of $m$ jobs in the queue.

\subsection{Designing a utility function}
Having estimations of the fidelity and queuing characteristics of a machine means that we are ready to design the utility function.
Utility function based scheduling has been utilized effectively in classical supercomputing~\cite{allcock2017experience}. 
First, maximizing the function should result in a job schedule that provides with a good balance between fidelity and queuing time.
Second, the function should also account for QOS requirements and the impact of calibrations and stale compilations on the utility of the machine.
Beyond the above (but not pursued in this work), the utility function could account for user priorities, improved machine utilization etc.
We use a balanced linear equation (sum of linear terms) of the form: $\Sigma (a_i*x_i)$, where $x_i$ is the feature (describing queuing time metric, machine-application fidelity metric, QOS satisfied or not, expected calibration crossover or not etc)  and $a_i$ is the coefficients (we currently only use -1 / 0 / 1), to design the utility function. 
More complex functions can be designed with suitably tuned coefficients if required.
Next, we discuss the how this utility function compares to the naive scheduling baselines.

\begin{figure*}[t]
     \centering
     \begin{subfigure}[b]{0.32\textwidth}
         \centering
         \includegraphics[width=\textwidth]{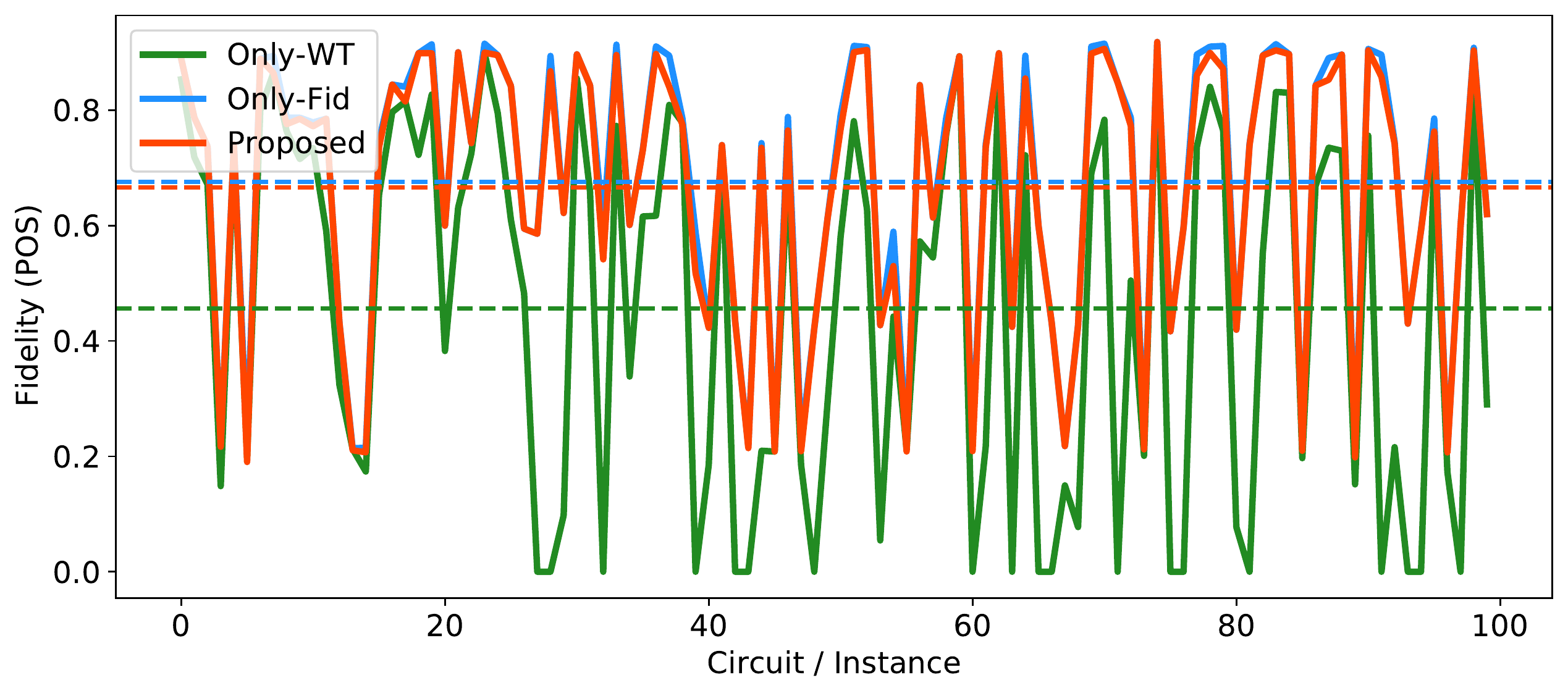}
         \caption{Fidelity (Low load)}
     \end{subfigure}
     \begin{subfigure}[b]{0.32\textwidth}
         \centering
         \includegraphics[width=\textwidth]{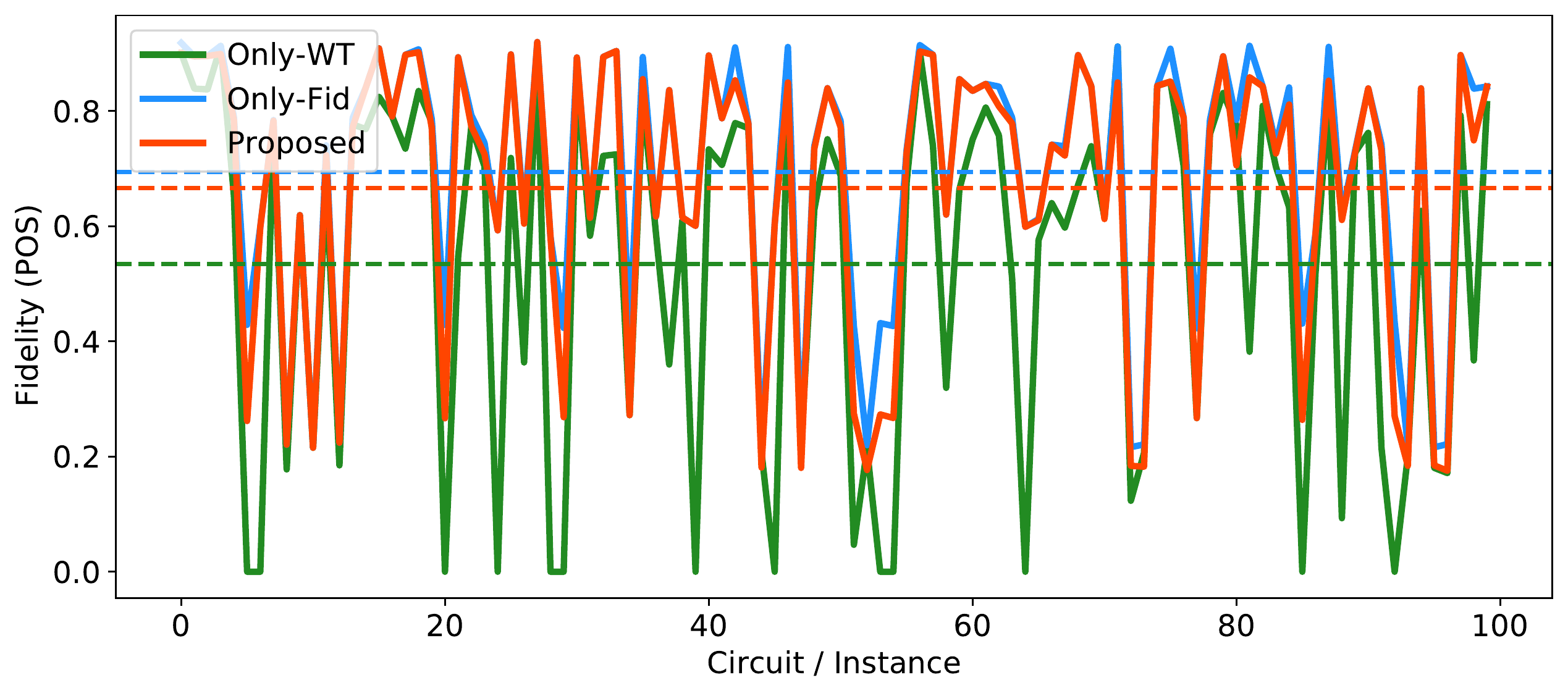}
         \caption{Fidelity (High load)}
     \end{subfigure}
     \begin{subfigure}[b]{0.32\textwidth}
         \centering
         \includegraphics[width=\textwidth]{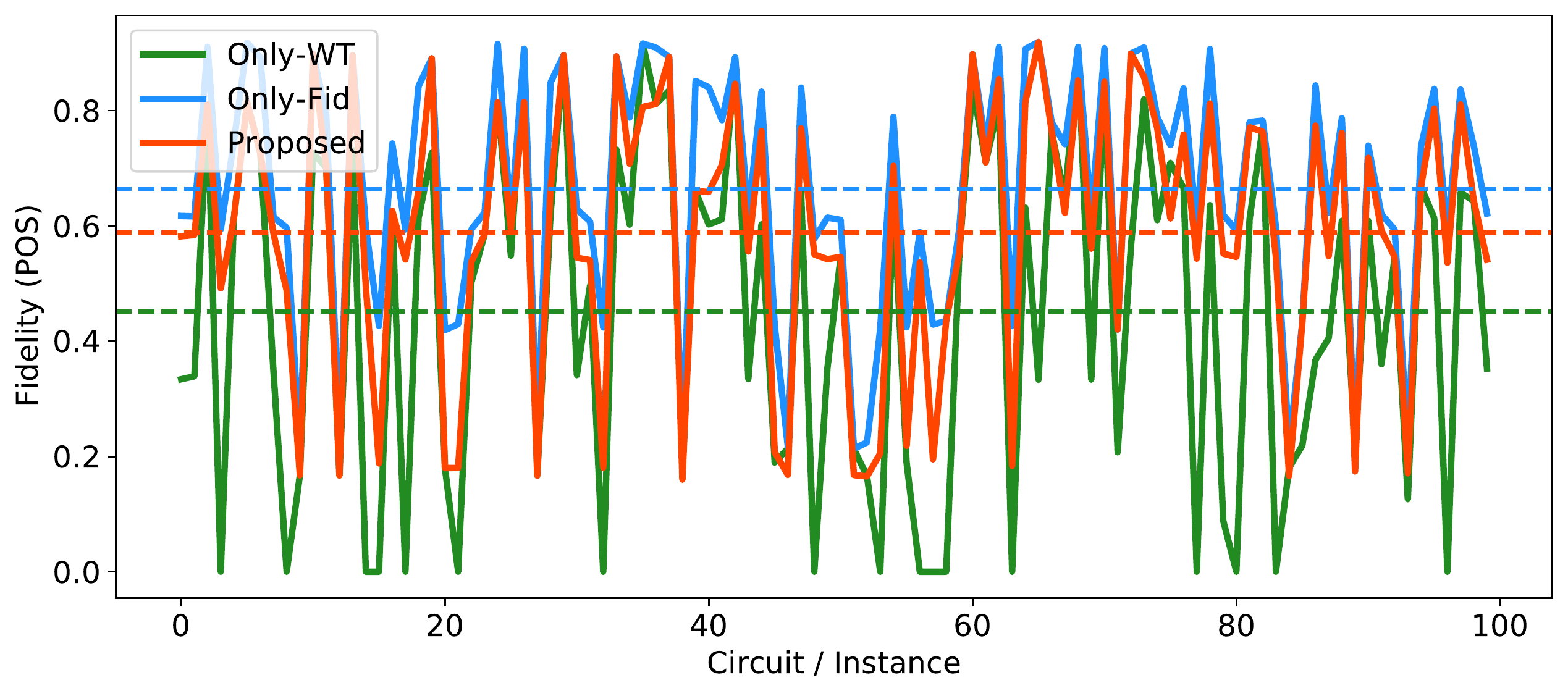}
         \caption{Fidelity (Random load)}
     \end{subfigure}
     \begin{subfigure}[b]{0.32\textwidth}
         \centering
         \includegraphics[width=\textwidth]{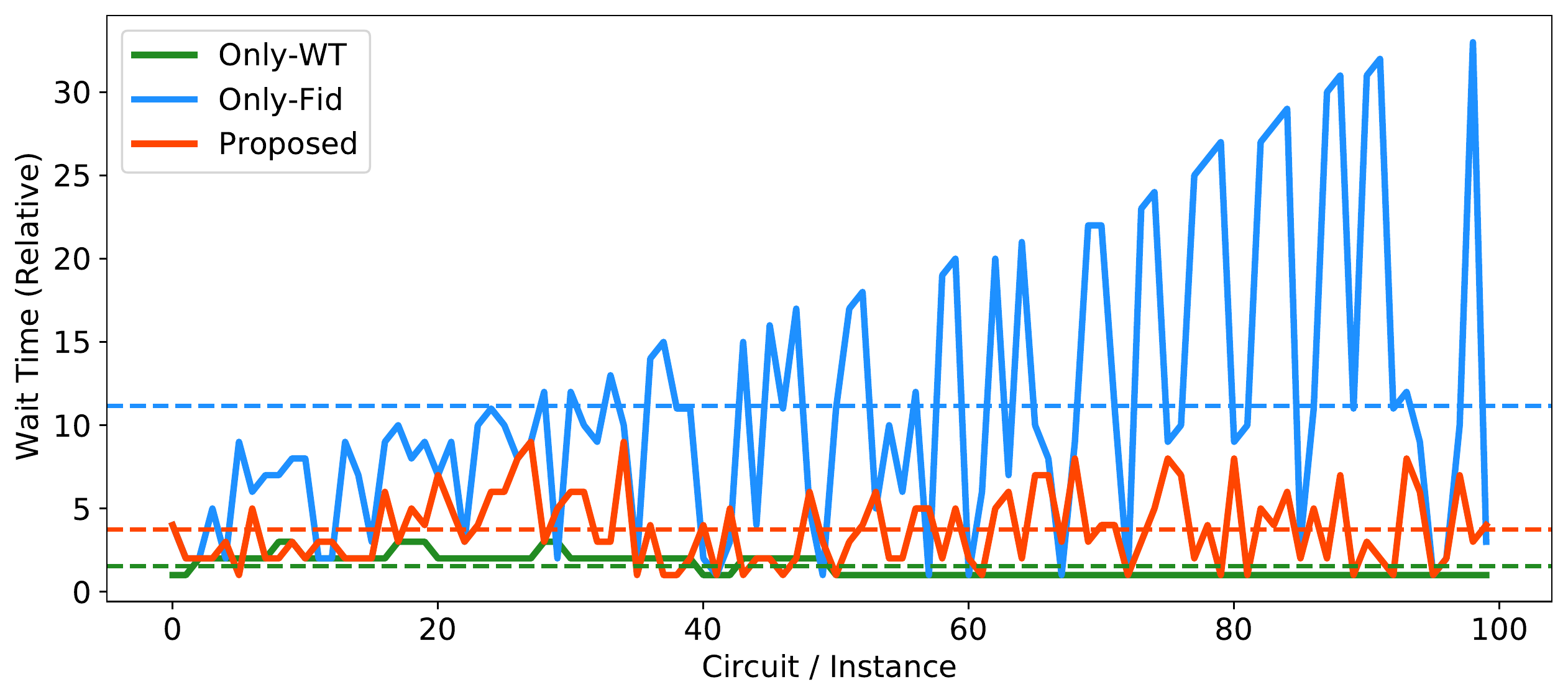}
         \caption{Wait Times (Low load)}
     \end{subfigure}
     \begin{subfigure}[b]{0.32\textwidth}
         \centering
         \includegraphics[width=\textwidth]{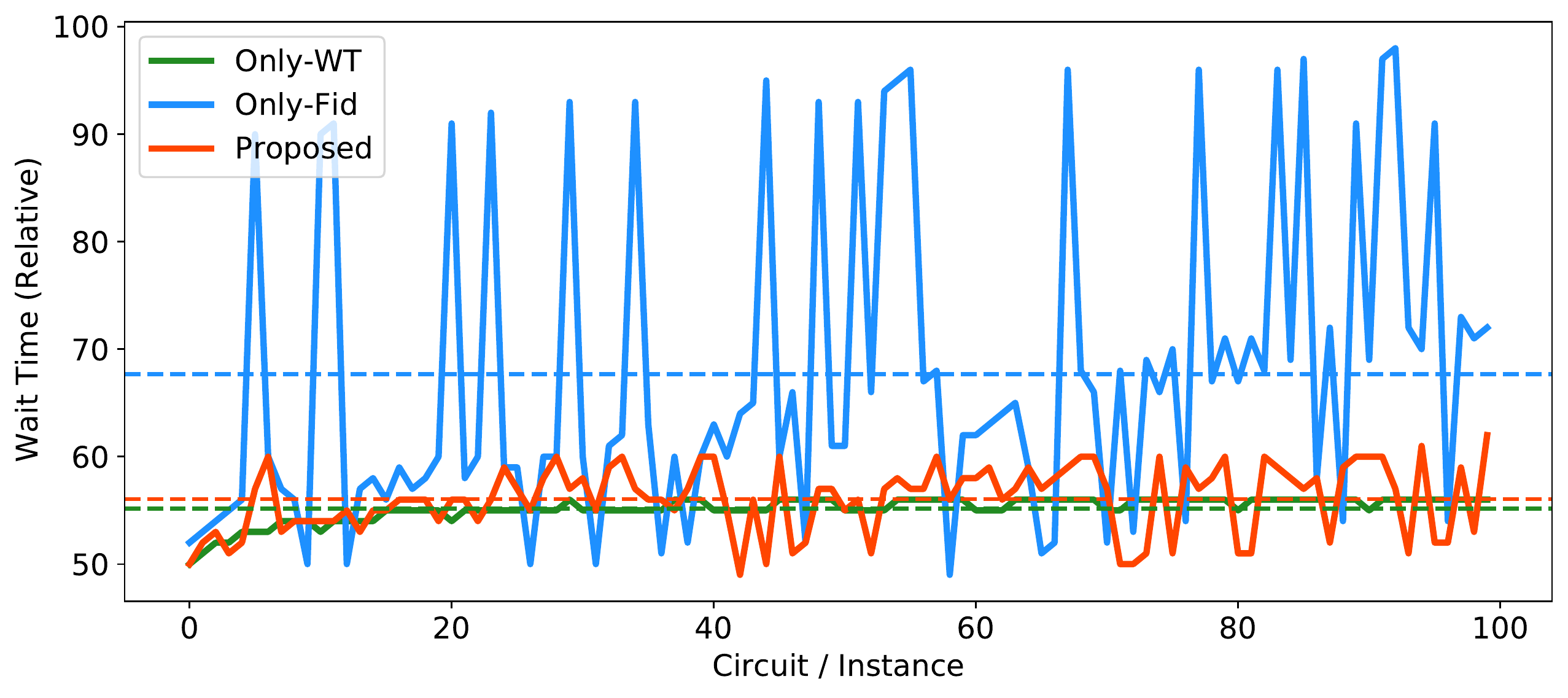}
         \caption{Wait Times (High load)}
     \end{subfigure}
     \begin{subfigure}[b]{0.32\textwidth}
         \centering
         \includegraphics[width=\textwidth]{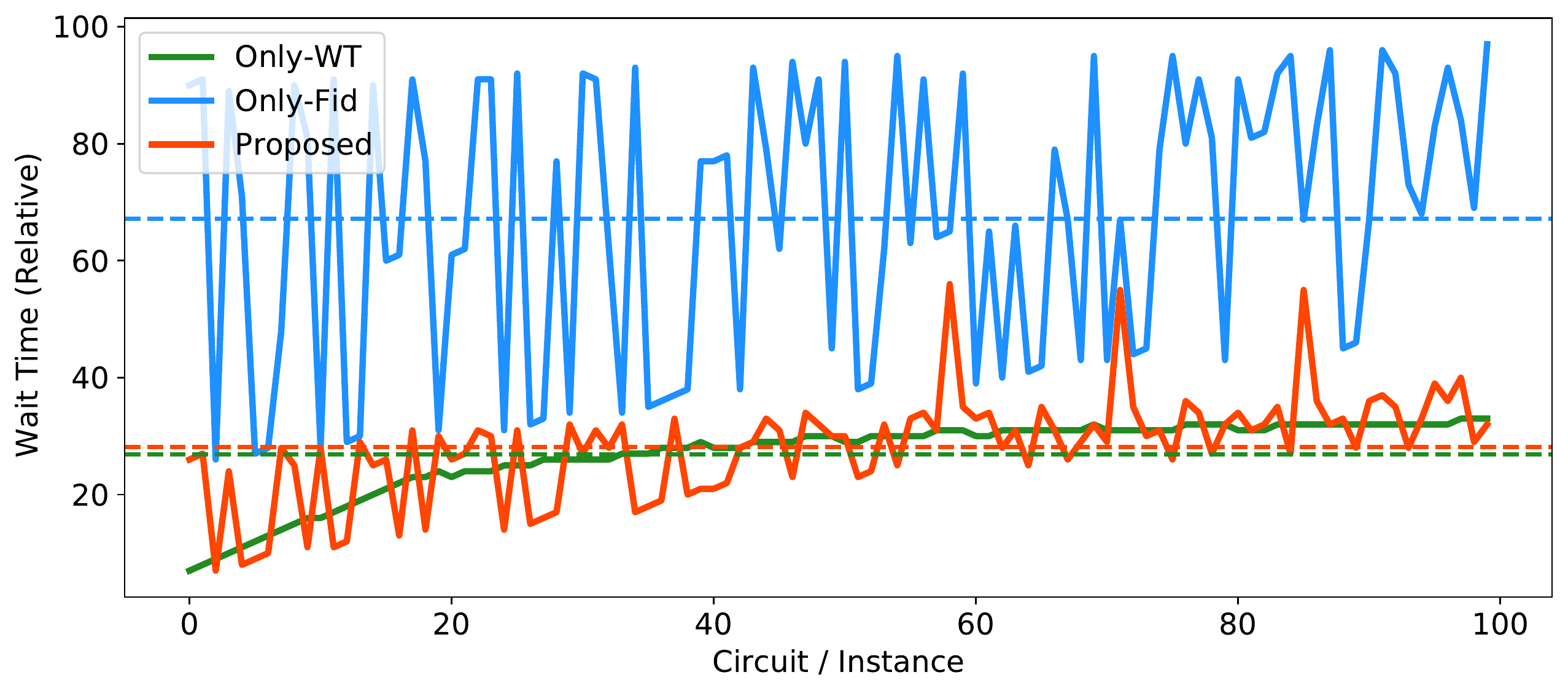}
         \caption{Wait Times (Random load)}
     \end{subfigure}
        \caption{Optimizing for Fidelity and Wait Times}
        \label{Fig:QCE21_Base}
\end{figure*}

\subsection{Balancing between fidelity and wait times}

Fig.\ref{Fig:QCE21_Base} shows comparisons of the effectiveness of the proposed approach (\emph{Proposed}) in balancing wait times and fidelity, in comparison to baselines which target only fidelity maximization (\emph{Only-Fid}) or only wait time reduction (\emph{Only-WT}).
The solid lines show per-instance metrics while the dashed lines so averages.
These comparisons are built by running the schedulers on a sequence of 100 circuits, which are picked randomly from our benchmark set, to be scheduled on our simulated quantum cloud system.
In this experiment we do use any QOS requirements nor do we look at the impact of calibration crossovers.

\textbf{Low Load:}
Fig.\ref{Fig:QCE21_Base}.a shows how fidelity varies across the sequence of jobs executed on our simulated quantum cloud system at low load.
At low load across machines, we would ideally want the highest fidelity machines to be chosen, since the queuing times are not significant and thus best results are worth the short wait.
Clearly, the fidelity achieved by the \emph{Only-Fid} is the highest as it always selects the machine which is predicted have the highest fidelity for application execution.
The fidelity achieved by \emph{Only-WT} is substantially lower, achieving only about 70\% of the \emph{Only-Fid} fidelity on average.
This is intuitive because \emph{Only-WT} simply selects machines which have the least queuing time.
On the other hand, our proposed approach is within 1\% of the ideal fidelity (\emph{Only-Fid}) and and roughly 40\% higher average fidelity compared to \emph{Only-WT}.
Fig.\ref{Fig:QCE21_Base}.d shows the wait times for this same low load usecase.
As expected the wait times of \emph{Only-WT} are always at the minimum - at load load, there are always relative free machines to execute jobs almost immediately.
\emph{Only-Fid} has considerably longer wait times even in this load load scenario, primarily because only a few high fidelity machines (like those to the right of Fig.\ref{Fig:QCE_Machine_Fid}) are being constantly targeted.
Our \emph{Proposed} approach shows higher wait times than the \emph{Only-WT} scenario but is still negligible at low load, while its wait time is roughly 3x lower on average (and up to 7x lower) than the \emph{Only-Fid} approach.
Clearly the proposed approach is not sacrificing on fidelity, but at the same time achieves reasonably low queuing times.
At low load, this is optimal for the system.

\textbf{High Load:}
Fig.\ref{Fig:QCE21_Base}.b shows how fidelity varies across a sequence of jobs executed on our simulated quantum cloud system at high load.
At high load across machines, we would ideally accept some loss in fidelity in order to achieve reasonable queuing times, though we would still want the fidelity to be substantial enough for realistic benefits.
First, Fig.\ref{Fig:QCE21_Base}.b shows that even at high load, our \emph{Proposed} approach's average fidelity is within 5\% of the fidelity-focused \emph{Only-Fid} approach but roughly 25\% better than the queuing focused \emph{Only-WT} approach.
Second, Fig.\ref{Fig:QCE21_Base}.e shows that the wait times of \emph{Proposed} very closely follows that of \emph{Only-WT}, which is ideal at high load, and is roughly 20\% lower than the \emph{Only-Fid} approach on average (up to 2x lower).
Clearly the proposed scheduler is not sacrificing on wait times, but at the same time achieves reasonably high fidelity.
At high load, this is optimal for the system.

\textbf{Random Load:}
Finally at random load, we see again that the \emph{Proposed} approach achieves 30\% higher fidelity than \emph{Only-WT} (Fig.\ref{Fig:QCE21_Base}.c) and 2.3x lower queuing times than \emph{Only-Fid} (Fig.\ref{Fig:QCE21_Base}.f), clearly highlighting the benefits of the proposed scheduler.
Note that the coefficients of the utility function can potentially be fine tuned so that these margins are even better.

\begin{figure*}[t]
     \centering
     \begin{subfigure}[b]{0.32\textwidth}
         \centering
         \includegraphics[width=\textwidth]{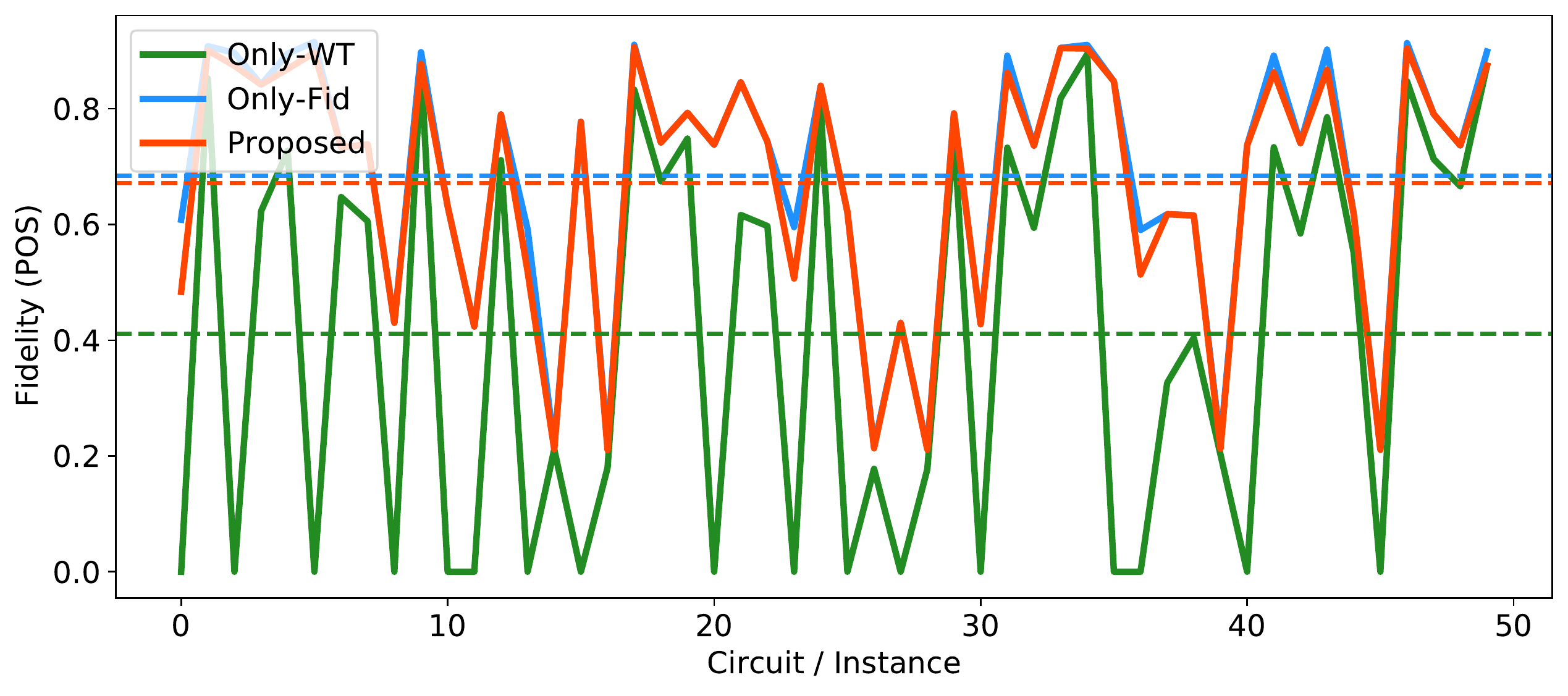}
         \caption{Fidelity (QOS 50)}
     \end{subfigure}
     \begin{subfigure}[b]{0.32\textwidth}
         \centering
         \includegraphics[width=\textwidth]{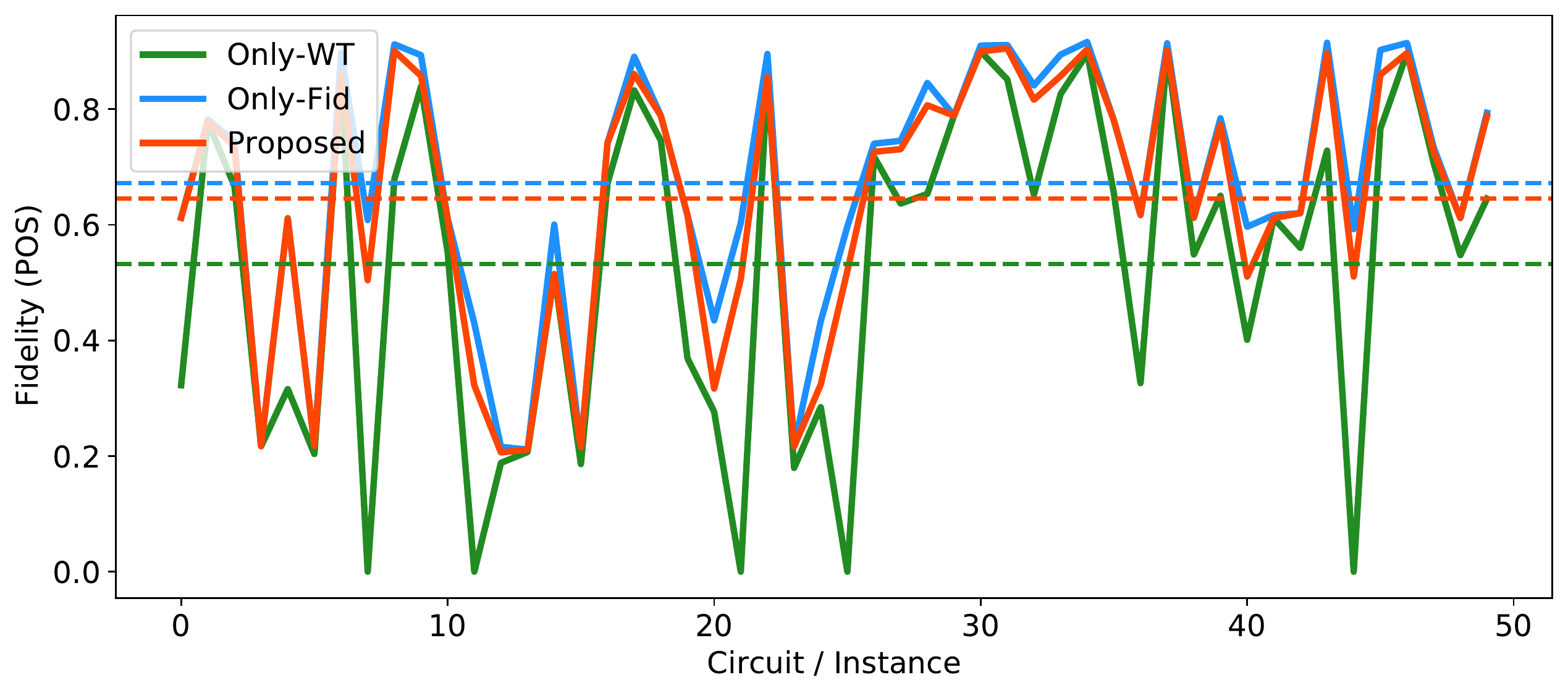}
         \caption{Fidelity (QOS 25)}
     \end{subfigure}
     \begin{subfigure}[b]{0.32\textwidth}
         \centering
         \includegraphics[width=\textwidth]{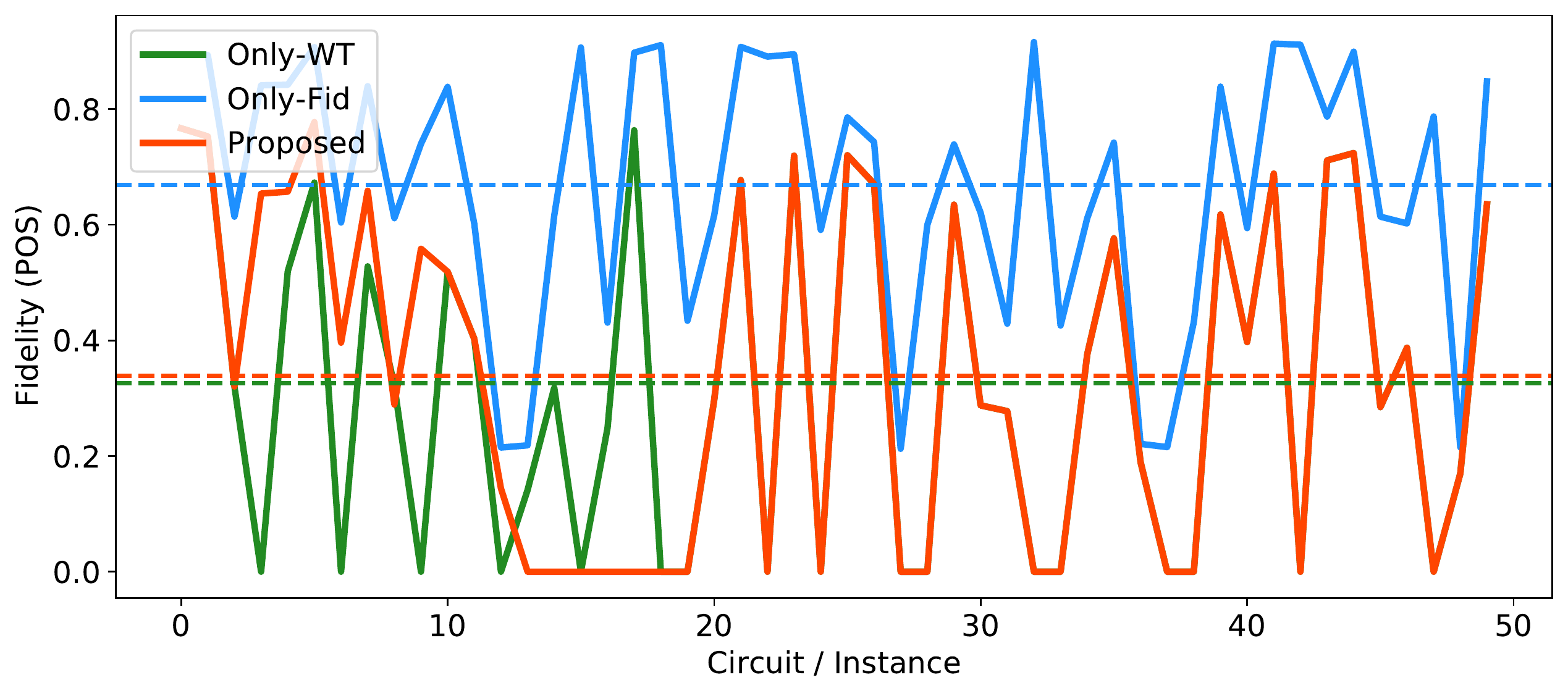}
         \caption{Fidelity (QOS 10)}
     \end{subfigure}
     \begin{subfigure}[b]{0.32\textwidth}
         \centering
         \includegraphics[width=\textwidth]{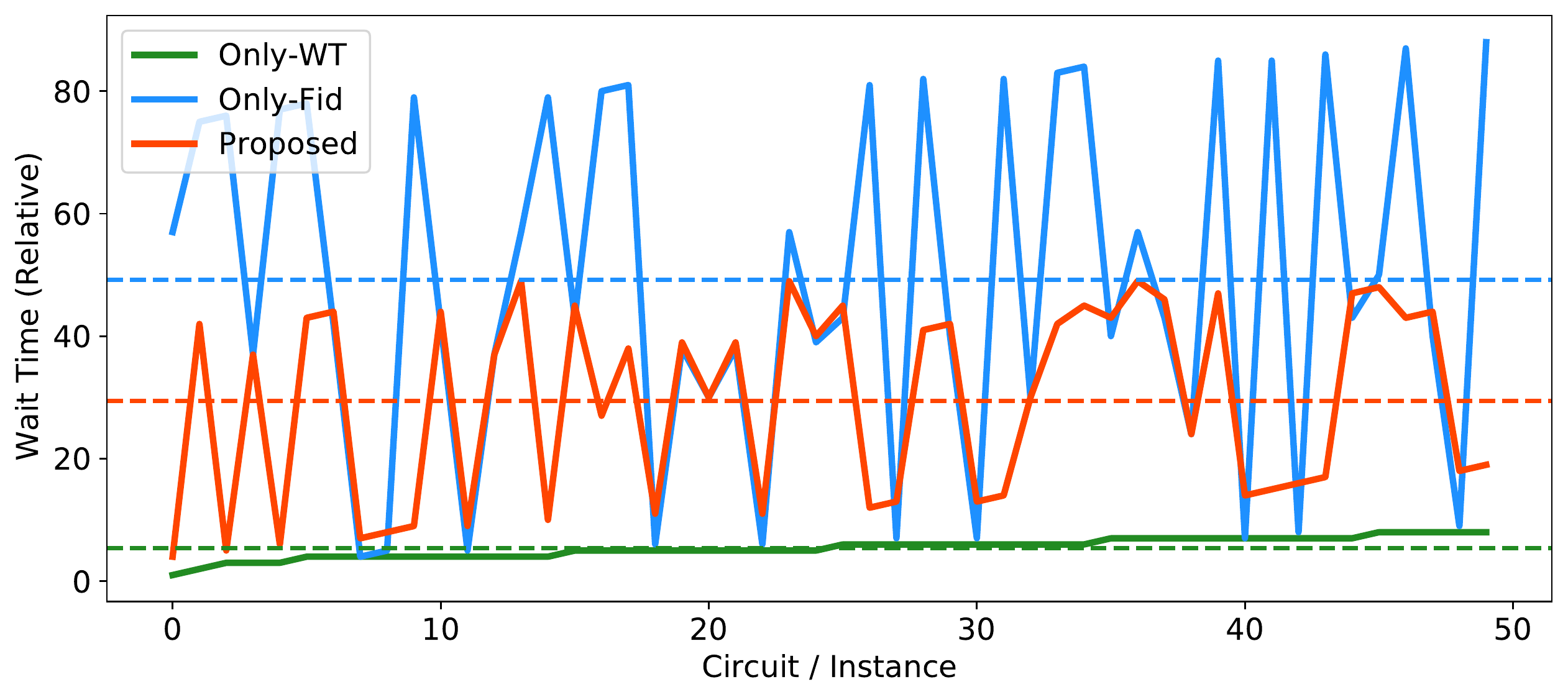}
         \caption{Wait Times (QOS 50)}
     \end{subfigure}
     \begin{subfigure}[b]{0.32\textwidth}
         \centering
         \includegraphics[width=\textwidth]{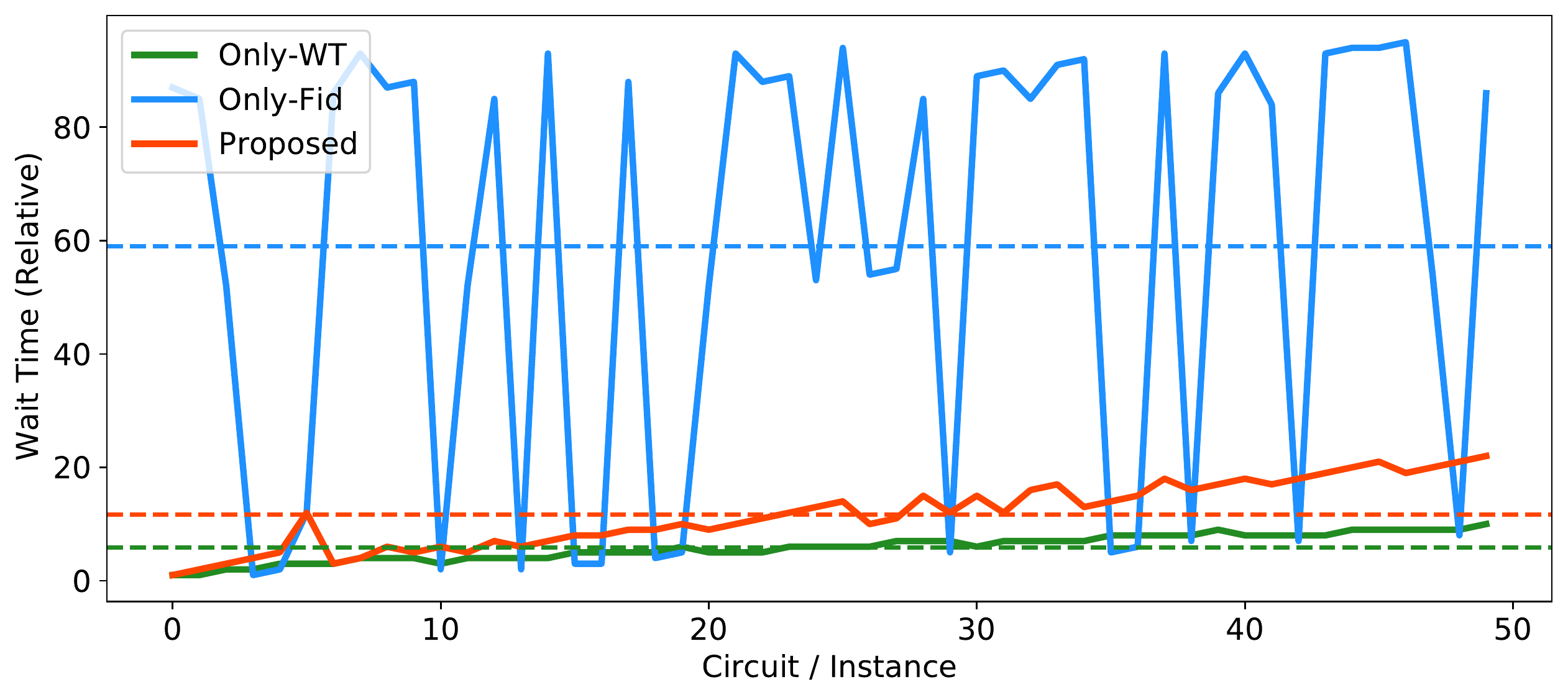}
         \caption{Wait Times (QOS 25)}
     \end{subfigure}
     \begin{subfigure}[b]{0.32\textwidth}
         \centering
         \includegraphics[width=\textwidth]{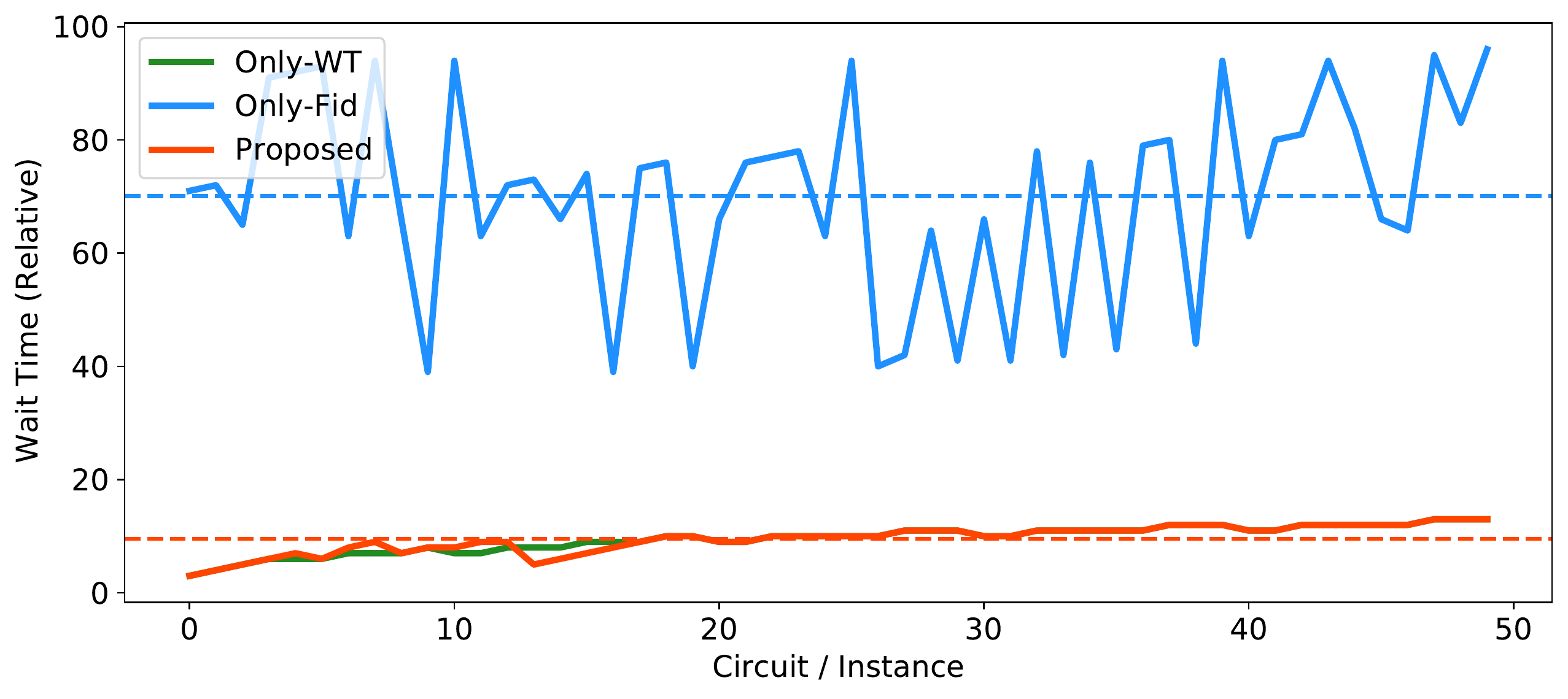}
         \caption{Wait Times (QOS 10)}
     \end{subfigure}
        \caption{Incorporating Quality of Service}
        \label{Fig:QCE21_QOS}
\end{figure*}

\subsection{Accounting for QOS specifications}
In Fig.\ref{Fig:QCE21_QOS} we perform the same analysis but with QOS specifications in terms of targeted maximum queuing times.

\textbf{QOS 50:}
Fig.\ref{Fig:QCE21_QOS}.a and Fig.\ref{Fig:QCE21_QOS}.d show the fidelity and wait times respectively for "QOS50" which means a wait time of up to "50" is tolerated by these jobs.
Clearly from Fig.\ref{Fig:QCE21_QOS}.a, the relaxed QOS requirements means that \emph{Proposed} is able to achieve nearly maximum fidelity, comparable to the \emph{Only-Fid} approach and 60\% better than that achieved by the \emph{Only-WT} approach.
Further, from Fig.\ref{Fig:QCE21_QOS}.d it is evident that the QOS requirements are always met by \emph{Proposed} unlike \emph{Only-Fid} which constantly overshoots it.

\textbf{QOS 25:}
Fig.\ref{Fig:QCE21_QOS}.b and Fig.\ref{Fig:QCE21_QOS}.e show the fidelity and wait times respectively for "QOS25" which means a wait time of up to "25" is tolerated by these jobs, a tighter bound.
In Fig.\ref{Fig:QCE21_QOS}.b, the stricter bound means that \emph{Proposed} sacrifices about 5\% of the maximum fidelity but still achieves 20\% higher fidelity than the \emph{Only-WT} approach.
Further,  from Fig.\ref{Fig:QCE21_QOS}.e it is evident that the QOS requirements are still met by \emph{Proposed}.

\textbf{QOS 10:}
Fig.\ref{Fig:QCE21_QOS}.c and Fig.\ref{Fig:QCE21_QOS}.f show the fidelity and wait times respectively for "QOS10" which means a wait time of up to "10" is tolerated by these jobs, a very strict requirement.
In Fig.\ref{Fig:QCE21_QOS}.c, this results in the fidelity of \emph{Proposed} falling down to match \emph{Only-WT} meaning that meeting such a strict QOS requires the scheduler to solely focus on wait time optimization.
Fig.\ref{Fig:QCE21_QOS}.f shows that \emph{Proposed} is still able to match the QOS bounds by mimicking the \emph{Only-WT} schedule.

Overall, it is clear that depending on the QOS specification, the proposed scheduler is able to adjust the job schedule to maximize fidelity while attempting to meet the QOS constraints, however strict they may be.

\begin{figure*}[t]
     \centering
     \begin{subfigure}[b]{0.32\textwidth}
         \centering
         \includegraphics[width=\textwidth]{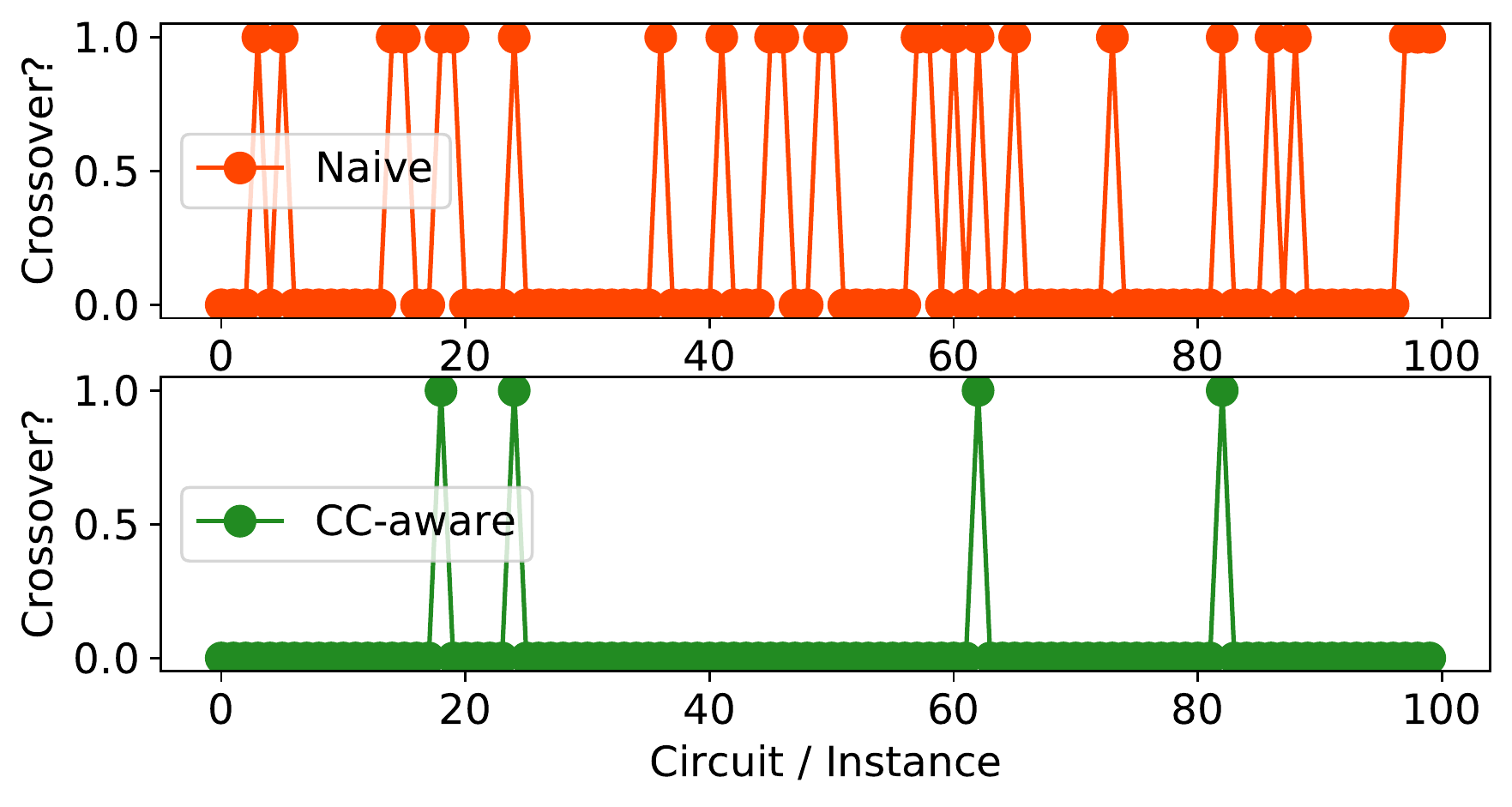}
         \caption{Low Load}
     \end{subfigure}
     \begin{subfigure}[b]{0.32\textwidth}
         \centering
         \includegraphics[width=\textwidth]{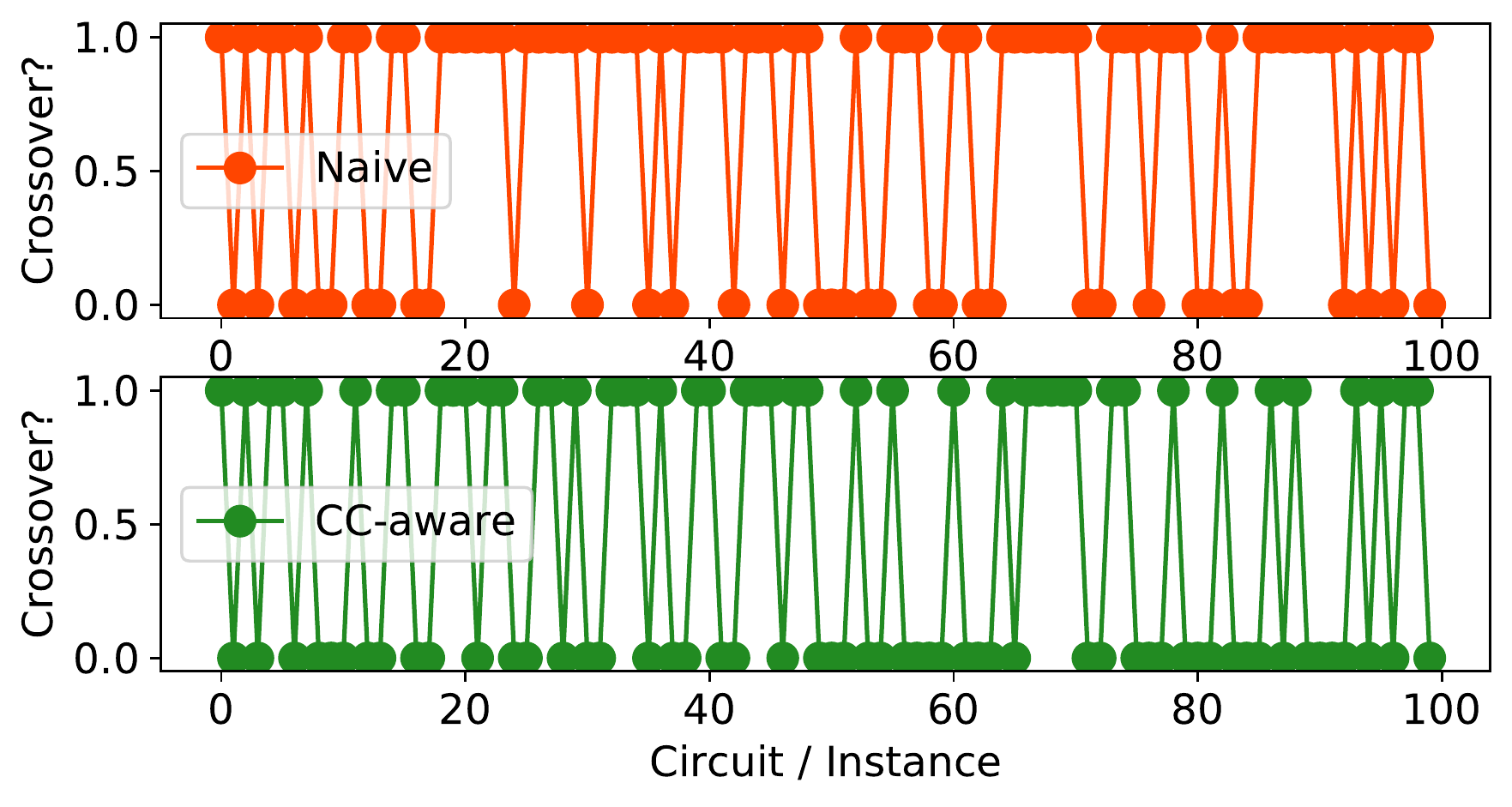}
         \caption{High Load}
     \end{subfigure}
     \begin{subfigure}[b]{0.32\textwidth}
         \centering
         \includegraphics[width=\textwidth]{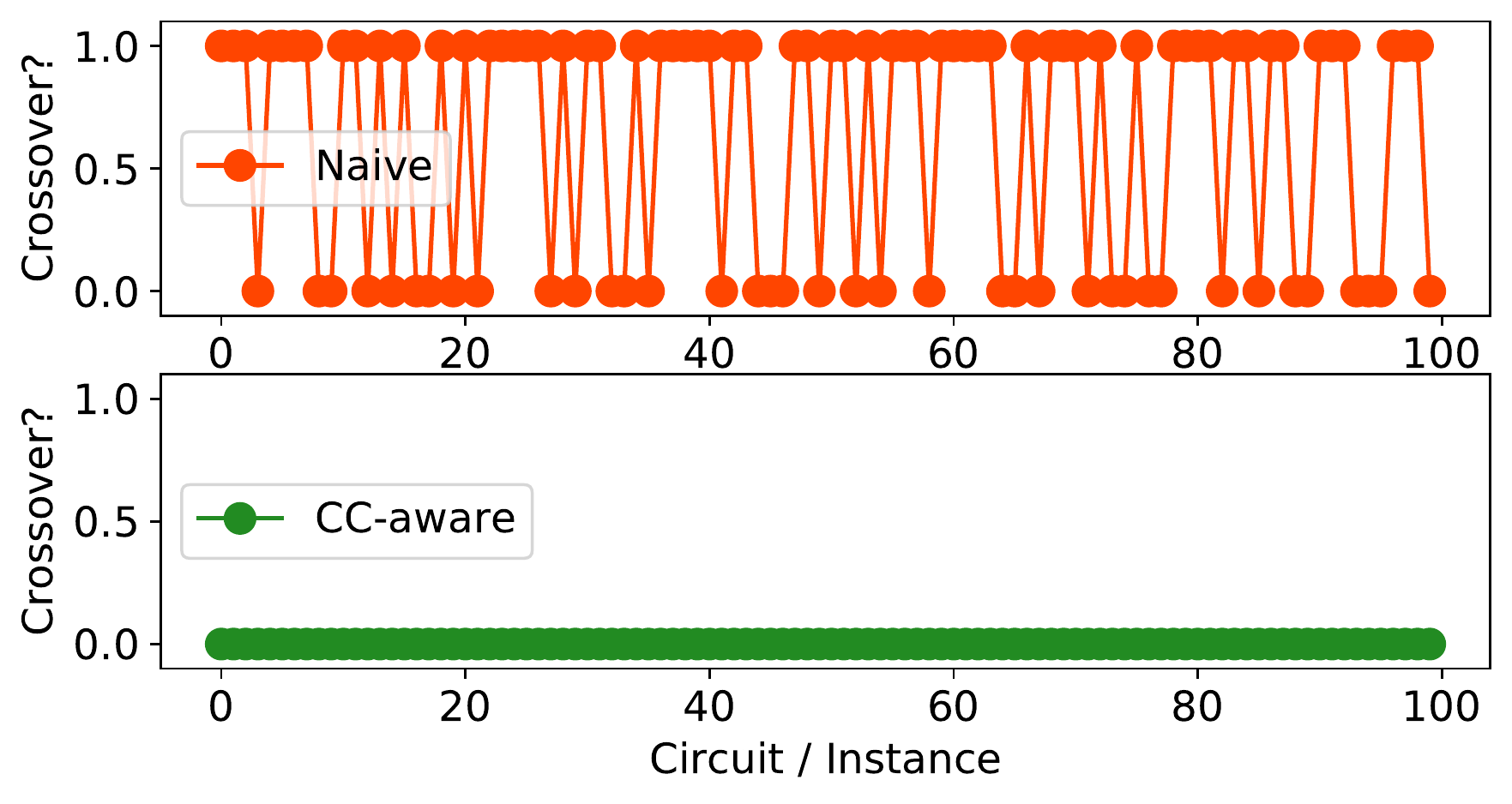}
         \caption{Staggered Calibration (High Load)}
     \end{subfigure}
        \caption{Accounting for Calibration Crossovers}
        \label{Fig:QCE21_CC}
\end{figure*}

\subsection{Avoiding Calibration Crossovers}
In Fig.\ref{Fig:QCE21_CC} we analyze the scheduler's capability to avoid calibration crossovers.
As discussed earlier, recalibration of machines (which, for example, is performed once a day around midnight by IBM Quantum machines) results in changes to error rates and device characteristics, meaning that machine-aware compilation on an old calibration cycle is not optimal for execution on a new calibration cycle.
This is especially critical in terms of our proposed scheduler since the scheduler estimates fidelity across the variety of machines based on information extracted post-compilation for each machine.

In Fig.\ref{Fig:QCE21_CC}.a shows a low load scenario.
On the y-axis, '1' implies a crossover occurred while '0' means no crossover.
A \emph{Naive} scheduling approach without accounting for calibration cycles results in multiple instances of crossover as shown in red.
On the other hand, accounting for calibration cycle information as part of the utility function adds a significant penalty to the  function if the crossover occurs, and thus is avoided whenever possible.
At low load, it is evident that a \emph{CC-aware} approach almost always avoids crossover.
Even if a job is scheduled late in the calibration cycle, the job is scheduler onto a machine with very low queuing time, almost always allowing the job to complete in the current cycle.

Unfortunately, avoiding crossovers is more challenging at high load.
Fig.\ref{Fig:QCE21_CC}.b shows that the \emph{CC-aware} approach still results in a large number of crossovers even though it is lower than the \emph{Naive approach}.
This is because, if a job is scheduled, say, 3 hours before the end of the calibration cycles of all machines, and if all the machines have queuing times greater than 3 hours, then it is impossible to avoid the crossover.

To overcome this, we instead propose a staggered calibration approach wherein machines are not calibrated all at nearly the same time (around midnight in North America), but instead the machine calibrations are distributed evenly throughout the day.
This means that irrespective of when a job is scheduled, there are always machines with considerable time left in their current calibration cycle, potentially allowing for one of those machines to be chosen for the job and thus having it complete execution within the current cycle on that machine.
Fig.\ref{Fig:QCE21_CC}.c shows the effect of this approach at high load.
The \emph{CC-aware} scheduling is now always able to avoid calibration crossovers, as is optimal for a machine calibration aware compilation approach.
Note that it is possible to design more intelligent staggered calibration policies based on observing queuing times on each machine, job arrival patterns etc.

\section{Limitations and Future Directions}

This work is a first step towards optimized engineering solutions for the quantum cloud.
It has certain limitations and opportunities for improvement, as discussed below:

    \circled{1}\  Note that the behaviors observed from the data shown here are a consequence of the usage policies established by IBM that govern machine behavior (such as fair-sharing policies) as well as the users themselves.
    Further, there are subsystems and interactions governing authentication, validation, diagnostics, etc. which are not considered.
    
    \circled{2}\  As presented, the work scheduler optimization does not consider other critical aspects of job scheduling like user priorities, improving machine utilization, drift in machine characteristics as well as dynamic changes to system load including machine reservations etc. They can be incorporated into the utility function model proposed in this work. 
    
    \circled{3}\  This work proposes compilation across multiple machines before choosing the right machine for execution.
    This might not scale well as applications become more complex and the number of machines increase. Thus it is important to identify machine execution characteristics (both application-independent and application-dependent) which can be estimated without compilation, which can then be used to shortlist the number of machines. This could include machine qubits, connectivity/topology, average machine-wide error rates etc.

    \circled{4}\  There is room to improve the fidelity correlation and execution time predictions model especially when targeting the complex characteristics of real quantum machine execution. More features can be added to the predictors, as well a more advanced learning model can be utilized.  

    \circled{5}\  To reduce the impact of machine calibration on job schedules, it is worth exploring more intelligent staggered calibration policies based on observing queuing times on each machine, job arrival patterns etc. Fine-grained calibrations such as IBM readout calibration can also be accounted for. 

   \circled{6}\ We envision evaluating in practice by testing on an actual cloud network and fine-tuning the functionality accordingly.

\section{Conclusion}
As quantum demand continuous to grow, it is imperative to efficiently manage quantum resources in the cloud.
This paper proposes to automate and improve scheduling quantum jobs to the quantum cloud.
It takes note of primary characteristics / requirements of quantum jobs and their scheduling, such as queuing times and fidelity trends across machines, as well as other aspects such as quality of service guarantees and machine calibration constraints. 
The proposed scheduler achieves a balance between application fidelity and queuing times while appreciating the quality of service requirements of users as well as the calibration cycles of the quantum machine.
Further it lays the foundation for more sophisticated quantum cloud job scheduling. 
In all, this work is a first step towards efficiently managing quantum machine usage, a precious commodity for the years to come.


\section*{Acknowledgement}
This work is funded in part by EPiQC, an NSF Expedition in Computing, under grants CCF-1730082/1730449; in part by STAQ under grant NSF Phy-1818914; in part by NSF Grant No. 2110860; in part by the US Department of Energy Office  of Advanced Scientific Computing Research, Accelerated Research for Quantum Computing Program; and in part by  NSF OMA-2016136 and in part based upon work supported by the U.S. Department of Energy, Office of Science, National Quantum Information Science Research Centers.  
GSR is supported as a Computing Innovation Fellow at the University of Chicago. This material is based upon work supported by the National Science Foundation under Grant \# 2030859 to the Computing Research Association for the CIFellows Project.
KNS is supported by IBM as a Postdoctoral Scholar at the University of Chicago and the Chicago Quantum Exchange.
FTC is Chief Scientist at Super.tech and an advisor to Quantum Circuits, Inc.

%
\IEEEpeerreviewmaketitle

\bibliographystyle{IEEEtranS}
\bibliography{refs}

\end{document}